\documentclass{jpp}
\usepackage{graphicx}

\usepackage[utf8]{inputenc}
\usepackage[T1]{fontenc}
\usepackage{amsmath}
\usepackage{bm}
\usepackage{subcaption}

\shorttitle{Impurity temperature screening in stellarators close to quasisymmetry}
\shortauthor{M.F. Martin, M. Landreman}

\title{Impurity temperature screening in stellarators close to quasisymmetry}

\author{Mike F. Martin
  \corresp{\email{mfmartin@umd.edu}}, Matt Landreman}

\affiliation{Institute for Research in Electronics and Applied Physics, University of Maryland, College Park, MD, 20742, USA}

\begin{document}

\maketitle

\begin{abstract}
    Impurity temperature screening is a favorable neoclassical phenomenon involving an outward radial flux of impurity ions from the core of fusion devices.
    Quasisymmetric magnetic fields are intrinsically ambipolar and give rise to temperature screening for low enough $\eta^{-1}\equiv d\ln n/d\ln T$.
    In contrast, neoclassical fluxes in generic stellarators will depend on the radial electric field, which is predicted to be inward for ion-root plasmas, potentially leading to impurity accumulation.
    Here we examine the impurity particle flux in a number of approximately quasisymmetric stellarator configurations and parameter regimes while varying the amount of symmetry-breaking in the magnetic field.
    Neoclassical fluxes have been obtained using the SFINCS drift-kinetic equation solver with the electrostatic potential $\Phi=\Phi(\psi)$, where $\psi$ is a flux surface label.
    Results indicate that achieving temperature screening is possible, but unlikely, at reactor-relevant conditions in the core.
    Thus, the small departures from symmetry in nominally quasisymmetric stellarators are large enough to significantly alter the neoclassical impurity transport.
    A further look at these fluxes when compared to a gyro-Bohm turbulence estimate suggests that neoclassical fluxes are small in optimized configurations compared to turbulent fluxes.
    Therefore, although neoclassical impurity accumulation is expected in most situations, the strength of the turbulence may render this irrelevant.
\end{abstract}

\section{Introduction}
The ideal makeup of particles in the core of magnetic confinement fusion devices would consist exclusively of particles participating in the fusion reaction.
The presence of any impurity ions can degrade fusion performance by means of fuel dilution and radiative cooling of the plasma \citep{Post61,W7AS}.
Removing impurities from the plasma core and preventing further accumulation is then of primary importance in present devices, and when designing future experiments.

Due to the symmetric nature of a tokamak, its neoclassical transport properties yield a distinct advantage over non-axisymmetric configurations because they are independent of the radial electric field, $E_r$, at leading order in $(\rho_i/L)$ \citep{Helander_book,Rutherford70}.
Here, $\rho_i$ is the ion gyroradius, and $L$ is some equilibrium scale length.
In particular, if certain conditions are met (see Section \ref{ambipolarity}), this absence of $E_r$ in the transport equations leads to a property known as temperature screening \citep{Wade00}, which guarantees an outward radial flux of impurities for large enough temperature gradients.

Conversely, unoptimized stellarator designs have been predicted to behave poorly with regards to impurity accumulation \citep{W7AS,Igitkhanov06,Ida09,Helander12}.
The lack of toroidal symmetry in the magnetic field complicates the transport quantities because of a dependence on $E_r$ in order to maintain ambipolarity of the constituent particle fluxes.
This can become an issue in reactor-relevant plasmas, which are expected to operate in the ion-root regime \citep{Maassberg99}, where the negative (inward) $E_r$ will tend to pull high-Z impurities into the core.
Here, $Z$ is the impurity ion charge.
Recent work \citep{Helander17}, however, has found that outward impurity fluxes can be achieved in the ``mixed-collisionality" regime in a stellarator, alleviating some of the concern.

Improving the behavior of impurities in stellarators could be addressed by contemporary stellarator design optimization, where one of the current foci is on quasisymmetric magnetic fields \citep{Nuhrenberg88}. Quasisymmetric fields have the allure of possessing the superior transport properties of tokamaks alongside the stability of stellarators.
Ideally, perfect quasisymmetry would lead to neoclassical and guiding-center transport properties identical to tokamaks \citep{Boozer83,Pytte81}.
However, it has been shown \citep{Garren91} that perfect quasisymmetry can likely be achieved only on a single flux surface.
Therefore, any future experiment or reactor will necessarily have some finite degree of symmetry-breaking.
This makes it important to study quasisymmetric equilibria with some departure from perfect symmetry.

In this paper, we examine the temperature screening effect using the SFINCS \citep{SFINCS} (Stellarator Fokker-Planck Iterative Neoclassical Conservative Solver) drift-kinetic solver to calculate the impurity particle flux for a number of quasisymmetric equilibria.
As we proceed, it will be necessary to distinguish between a perfectly quasisymmetric magnetic field, and the quasisymmetry of configurations such as the National Compact Stellarator Experiment (NCSX) \citep{NCSX} or the Helically Symmetric Experiment (HSX) \citep{HSX}.
For example, the magnetic field of HSX is quasisymmetric in the sense that its quasisymmetric harmonics are dominant compared to the smaller, but non-zero, symmetry-breaking harmonics.
Such a magnetic field will be referred to as the actual, or true, magnetic field of a configuration.
A \textit{perfectly} quasisymmetric magnetic field is one in which the symmetry-breaking modes are identically zero.

With this distinction, the unanswered question we would like to address is whether, in a nominally quasisymmetric stellarator with realistic deviations from perfect symmetry, the sign of the neoclassical impurity flux is outward like in tokamaks, or inward like in a generic stellarator.

By altering the magnitude of symmetry-breaking harmonics in the magnetic field of a given equilibrium (see Section \ref{bmn_scale}), we are able to probe the region where temperature screening is lost.
Holding the temperature constant, this effect was studied at 3 distinct densities, and correspondingly 3 distinct collisionalities.
At the lowest collisionality, no configurations are able to maintain an outward impurity flux at the true magnetic field, even for $\eta^{-1}\equiv d\ln(n_a)/d\ln(T_a)=0$, where $a$ refers to species.
(Introducing a finite peaked density gradient for the main ions, $\eta_i^{-1}>0$, always makes the impurity particle flux more inward, which is explained in Section \ref{finite_eta}).
Increasing the collisionality has a favorable effect, where some configurations were even found to have an outward impurity flux.
However, there is an upper collisionality limit, beyond which temperature screening is not observed for most configurations, even in perfect quasisymmetry.

Impurity accumulation in perfect quasisymmetry with $\eta^{-1}=0$ can either be caused by exceeding the collisionality limits where temperature screening is valid, or by a dependence of the neoclassical transport on $E_r$, indicative of a breakdown in the intrinsic ambipolarity assumption.
In the latter case, the $\bm{E}\times\bm{B}$ drift, $v_{E}$, is close to being in violation of the $v_E\sim\rho_* v_{t\alpha}$ ordering in deriving the equations solved in neoclassical codes.
In Section \ref{er_res}, we examine this in further detail and calculate the resonant radial electric field, $E_r^{res}$, in quasisymmetric configurations. One finds that $E_r^{res}$ is fundamentally smaller in quasi-axisymmetry (QA) than in quasi-helical symmetry (QH).

We have also compared the magnitude of the resulting neoclassical fluxes to a gyro-Bohm estimate for turbulence.
At reactor-relevant parameters, the neoclassical impurity particle flux did not exceed the respective turbulent flux for any impurity species or configuration.
Even in the presence of a strongly peaked $(\eta^{-1}=0.5)$ density gradient, in most configurations the neoclassical impurity particle flux is less than 10\% of the estimated turbulent value.
This suggests that regardless of whether a configuration can achieve temperature screening, the nature of the turbulence will determine the sign of the particle flux on a surface.

The total (bulk ion + impurity) neoclassical heat flux also did not exceed the turbulent contribution.
However, the ratio was larger than the analogous impurity particle flux ratio.
Furthermore, the neoclassical contribution is largest near the magnetic axis, and results indicate that turbulence becomes increasingly more dominant as one moves further out radially.
This is in agreement with experimental observations \citep{Pablant18,Canik07} in Wendelstein 7-X (W7-X) \citep{W7X} and HSX, which find that neoclassical transport is the dominant radial transport channel near the magnetic axis.

Finally, we compared the critical amount of symmetry-breaking that it takes to change the sign of the particle flux, $\epsilon_{sb}^c$, to two metrics that have been used to quantify symmetry on a flux surface.
These metrics are the effective helical ripple, $\epsilon_{eff}$, which is a measure of neoclassical transport in the $1/\nu$ regime, and the magnitude of the symmetry-breaking terms on a flux surface, $S$ (see Eq. \ref{eq:s}).
While it was found that there was some anti-correlation between $\epsilon_{sb}^c$ and $S$, there does not appear to be much of a relationship between $\epsilon_{eff}$ and $\epsilon_{sb}^c$. (This should not be surprising, however, if one considers that W7-X has a very low $\epsilon_{eff}$, yet it is far from quasisymmetry).
This difference between how $\epsilon_{eff}$ and $S$ depend on $\epsilon_{sb}^c$, a quantity related to symmetry, motivates a comparison between $\epsilon_{eff}$ and $S$.
Results indicate a configuration-specific dependence of $\epsilon_{eff}$ on $S$, which in many cases is non-monotonic.
There is thus a disconnect between these two quantities, such that minimizing the amount of symmetry-breaking on a flux surface does not simultaneously minimize $\epsilon_{eff}$.
So while $\epsilon_{eff}$ is a useful proxy for optimizing neoclassical transport in stellarator optimization, it is a poor proxy for achieving good quasisymmetric surfaces.

This paper is organized as follows: In Section \ref{drift_kinetic}, we introduce the governing equation and ordering assumptions of SFINCS in the results presented herein.
In Section \ref{ambipolarity}, we explain the principle of ambipolarity, the fundamentals of the temperature screening phenomenon, and the issues that arise in non-axisymmetric geometries.
In Section \ref{bmn_scale} we explain our approach to varying the degree of quasisymmetry on a flux surface.
The quasisymmetric configurations that have been explored, and the way that these equilibria have been scaled can be found in Section \ref{configs}.
Section \ref{er_res} explains an issue in present neoclassical stellarator codes based on the $v_E\sim\rho_* v_{ta}$ ordering, which limits the value of the radial electric field when impurities are included.
Section \ref{eps_eff} presents results on how the amount of symmetry-breaking, collisionality, and density gradient affect the behavior of the impurity particle flux for various quasisymmetric configurations.
Section \ref{turbulence} compares the results of Section \ref{eps_eff} to a gyro-Bohm estimate of turbulent particle and heat fluxes as a function of the impurity species, and normalized radius.
Finally, Section \ref{ripple} compares the effective helical ripple to the amplitude of symmetry-breaking terms on a flux surface.

\section{Background}
\subsection{Drift Kinetic Equation}\label{drift_kinetic}

Neoclassical transport follows from a drift-ordering of the Fokker-Planck equation in toroidal magnetic geometry, and solving the resulting drift-kinetic equation (Eq.19 in \citep{DKE_Hazeltine}).
The drift ordering assumes $\rho_{*a}=\rho_a/L\ll 1$, $v_E\sim\rho_{*a}v_{ta}$, $\partial/\partial t\sim\rho_{*a}^2v_{ta}/L$, and $\nu_a\sim v_{ta}/L$, where $\nu_a$ is the collision frequency.
The gyroradius of species $a$ is $\rho_a=v_{ta}/\Omega_a$, the thermal velocity $v_{ta}=\sqrt{2T_a/m_a}$, with $T_a$ and $m_a$ the temperature and mass of species $a$, respectively. The gyrofrequency is $\Omega_a=Z_aB/m_a c$, where $Z_a$ is the species charge, $B$ is the magnetic field magnitude, and $c$ is the speed of light.

Results in this paper have been obtained by solving the drift-kinetic equation (DKE) using the SFINCS \citep{SFINCS} code over a range of collisionality regimes, for various impurity ions.
SFINCS is a radially-local DKE-solver that has been generalized to non-axisymmetry, allowing for an arbitrary number of species, fully linearized Fokker-Planck collision operator, and the capability of simulating on-surface variations in the electrostatic potential, $\Phi_1$.
The exact form of the DKE that is solved in SFINCS for this paper (given by Eq.(16) in \citep{SFINCS}) is
\begin{equation}
\begin{split}
    \dot{\bm{r}}\cdot\left(\nabla f_{a1}\right)_{x_a,\xi} + \dot{x}_a\left(\frac{\partial f_{a1}}{\partial x_a}\right)_{\bm{r},\xi} + \dot{\xi}_a\left(\frac{\partial f_{a1}}{\partial\xi}\right)_{\bm{r},x_a} - C_a = -\left(\bm{v}_{ma}\cdot\nabla r\right)\left(\frac{\partial F_a}{\partial r}\right)_{W_{a0}},
    \label{eq:dke}
\end{split}
\end{equation}
where $F_a$ and $f_{a1}$ represent a Maxwellian distribution and the first-order perturbation to the distribution function, respectively.
The position vector is given by $\bm{r}$, the pitch angle is $\xi\equiv v_{\parallel}/v$, the velocity is $x_a\equiv v/v_{ta}$, $W_{a0}=v^2/2+Z_ae\Phi_0/m_a$ is the lowest-order total energy, $C_a$ is the collision operator, and $\bm{v}_{ma}$ is the magnetic drift velocity defined by
\begin{equation}
    \bm{v}_{ma}\cdot\nabla r = \frac{m_ac}{Z_aeB^2}\left(v_{\parallel}^2+\frac{v_{\perp}^2}{2}\right)\bm{b}\times\nabla B\cdot\nabla r.
\end{equation}
The coordinate $r=\sqrt{2\psi_t/B_{av}}$ is a surface label, where $2\pi\psi_t$ is the toroidal flux, and $B_{av}$ is some averaged magnetic field, such as the field on the magnetic axis.
The time derivative terms, $\dot{\bm{r}}$, $\dot{x}_a$, and $\dot{\xi}_a$ are the phase space particle trajectories.
Since SFINCS offers variations in how these trajectories are defined, we have chosen to use the ``full trajectories" definition (Eq.(17) in \citep{SFINCS}).
This choice takes into account the change in potential energy as a particle drifts radially, with the corresponding change to $\dot{\xi}_a$ in order to conserve the magnetic moment, $\mu$.
Finally, due to the linearization of the DKE, $\Phi_1$ effects are not present in Eq \ref{eq:dke}, and can only be calculated through the first-order quasineutrality condition.

When considering $\Phi_1$ in neoclassical transport, recent results \citep{Garcia-Reg17} indicate that it has only a moderate impact on the particle flux for highly-charged impurities, in the case of the \textit{non}-quasisymmetric Wendelstein 7-X (W7-X) stellarator. While not quasisymmetric, W7-X is still a neoclassically optimized stellarator and will have reduced radial excursions of helically-trapped particles, limiting the size of density variations on a flux surface.
Since $\Phi_1$ is closely connected to these density fluctuations \citep{Mynick84}, it is sensible to assume that $\Phi_1$ would not have a large impact on neoclassical transport.
In quasisymmetric experiments not deviating too far from perfect symmetry, it is reasonable to expect similar behavior.
For the simulations that follow, we will thus neglect any flux surface variations of the electrostatic potential and take $\Phi=\Phi_0(r)$.

\subsection{Ambipolarity and temperature screening}
\label{ambipolarity}

The property of ambipolarity can be expressed as
\begin{equation}
    \sum_a Z_a\Gamma_a = 0,
    \label{eq:ambipolarity}
\end{equation}
where $\Gamma_a$ is the radial component of the particle flux of species $a$
\begin{equation}
    \Gamma_a = \left\langle\int f_a\left(\bm{v}_{ma}\cdot\nabla r\right)\mathrm{d}^3 v\right\rangle,
\end{equation}
with $\left\langle\dots\right\rangle$ representing a flux surface-average.
This results from the charge density being small for length scales much longer than the Debye length.
Ambipolarity is then a statement that the flux surface-averaged radial current vanishes on each flux surface. 
While this is true in both tokamaks and stellarators, the radial electric field is set by different physical mechanisms, and $E_r$ only affects the ambipolarity condition in stellarators.
The value of the radial electric field that satisfies the ambipolarity condition in a non-axisymmetric plasma is referred to as the ambipolar radial electric field.

Neoclassical fluxes are determined from a linear combination of the equilibrium gradients in the system.
The radial neoclassical impurity particle flux can be written in the form of Eq (1) in \citep{Velasco17}:
\begin{equation}
    \Gamma_z=-n_z\sum_aL_{11}^a\left[\frac{1}{n_a}\frac{dn_a}{dr}-\frac{Z_aeE_r}{T_a}+\delta_a\frac{1}{T_a}\frac{dT_a}{dr}\right],
    \label{eq:gamma_z}
\end{equation}
where $r$ is an arbitrary radial coordinate, and $L_{11}^a$ and $\delta_a$ are transport coefficients \citep{Beidler11,Maassberg99,Velasco17} that can have a complicated dependence on $E_r$ and the collision frequency $\nu_z=\nu_{zi}+\nu_{zz}$, where
\begin{equation}
    \nu_{ab} = \frac{4\sqrt{2\pi}n_bZ_a^2Z_b^2\ln\Lambda}{3\sqrt{m_a}T_a^{3/2}}.
\end{equation}
In the case of a tokamak, the toroidal symmetry causes the electric field dependence to cancel out in Eq \ref{eq:ambipolarity}, making transport \textit{intrinsically ambipolar}.
This, in principle, allows for rapid toroidal rotation of the plasma, as the radial electric field profile is not governed by the ambipolarity constraint, but rather by angular momentum conservation \citep{Shaing86,Hirshman78,Helander_book,Abel13}.
Moreover, for a low-enough collisionality plasma with $\delta_i+\delta_z>0$, this cancellation assures a radially outward flux of the impurity species when $\eta^{-1}\equiv\nabla\ln n/\nabla\ln T < \eta_c^{-1}$, where $\eta_c^{-1}$ is some critical ratio of the density and temperature gradients.
This beneficial phenomenon is generally referred to as temperature screening.

The situation is less positive in stellarator geometries since the ambipolarity condition is dependent on the radial electric field, and the temperature screening effect is no longer guaranteed.
For fusion-relevant, high-density plasmas, the radial electric field is directed inward \citep{Maassberg99} and will presumably act to drive higher-$Z$ impurities into the core.
It should be stated clearly here that temperature screening is a \textit{neoclassical} effect, which is based on the intrinsic ambipolarity of turbulent fluxes even in non-axisymmetric geometries (to lowest order) \citep{Helander14}.
This permits a study of temperature screening without considering the effect of turbulence on the transport.

A potential solution to this situation in stellarators lies in the design of quasisymmetric configurations.
A truly quasisymmetric device, whose magnetic field varies through a fixed linear combination of Boozer angles on a flux surface, will have neoclassical and guiding-center transport properties identical to a tokamak, up to $O(\rho_{*a})$ \citep{Boozer83,Pytte81}.
However, evidence suggests that in the absence of axisymmetry, quasisymmetry cannot be achieved exactly throughout a volume \citep{Garren91}, meaning that quasisymmetric devices will necessarily deviate from symmetry to some level.
Therefore, it would be informative to optimization efforts to know how much breaking in the symmetry of the magnetic field can be tolerated before the temperature screening effect is lost.
In the following sections, we explore the effect that magnetic field symmetry-breaking has on the impurity particle flux.

\section{Magnetic field symmetry-breaking}
\label{bmn_scale}

Any magnetic field within a flux surface can be written as a sum of harmonics in the Boozer poloidal $\theta$, and toroidal $\zeta$ angles \citep{Boozer_coord}
\begin{equation}
    B(r,\theta,\zeta) = \sum_{m,n}B_{mn}(r)\mathrm{e}^{i(m\theta-n\zeta)}.
    \label{eq:b_generic}
\end{equation}
Only by expressing the magnetic field in Boozer coordinates will the property of quasisymmetry become apparent.
A magnetic field is considered quasisymmetric if its magnitude varies within a flux surface only through the fixed linear combination $\chi = M\theta+N\zeta$, where $M,N$ are fixed integers, one of which may be zero.
However, since perfect quasisymmetry is not practically achievable, it is possible to express the magnetic field of a quasisymmetric configuration as a sum of quasisymmetric and \textit{non}-quasisymmetric Boozer harmonics, the latter of which will be referred to as symmetry-breaking terms.
Therefore, symmetry-breaking terms with smaller magnitudes will produce better approximations to perfect symmetry.

Our approach to understanding temperature screening in stellarators exploits this fact by allowing one to adjust the amplitude of the symmetry-breaking terms by an overall, constant scaling factor.
The way we have decided to approach this is to expand in the harmonics of $1/B^2$, which can be expressed as

\begin{equation}
    \frac{1}{B^2} = \sum_q h_q(r)\mathrm{e}^{iq\chi} + \epsilon_{sb}\sum_{m,n}h_{mn}(r)\mathrm{e}^{i(m\theta-n\zeta)},
\end{equation}
where the quantities $h_q$ and $h_{mn}$ are the quasisymmetric and non-quasisymmetric harmonics of the $1/B^2$ expansion, respectively.
The parameter $\epsilon_{sb}$ is a scaling factor (fixed for a given simulation) that controls the amplitude of the symmetry-breaking terms.
The special case of $\epsilon_{sb}=0$ denotes a truly quasisymmetric field, while $\epsilon_{sb}=1$ corresponds to the original magnetic field that one would get from an equilibrium code.
By running simulations with $\epsilon_{sb}$ between these values, one can gain further insight into how temperature screening is affected under magnetic fields with varying degrees of symmetry-breaking. 

In the context of MHD, artificially scaling the magnetic field with $\epsilon_{sb}\ne 1$ will lead to a plasma that no longer satisfies the equations of an MHD equilibrium.
However, if we consider the work of Garren/Boozer \citep{Garren91}, it is likely that the construction of a \textit{single} quasisymmetric flux surface is possible.
Then, an arbitrarily quasisymmetric magnetic field could be constructed on one of many flux surfaces that, in principle, will satisfy an MHD equilibrium.
Since this applies to only one flux surface, our approach prevents the scaling of multiple flux surfaces simultaneously.

To understand our choice of expanding $1/B^2$, it is important to recognize that artificially scaling the magnetic field of an MHD equilibrium can potentially become problematic if large currents are introduced near rational surfaces \citep{Boozer81,Helander14}. 
This can be seen in the expression for the parallel current \citep{Helander14}
\begin{equation}
    J_{\parallel}\sim\frac{h_{mn}}{r-r_{mn}}\frac{dp}{dr}.
\end{equation}
Since our simulations will always assume a finite pressure gradient, the $h_{mn}$ modes must vanish on rational surfaces to avoid an infinite Pfirsch-Schl\"uter current.
Therefore, scaling $h_{mn}$ modes as opposed to $B_{mn}$ modes will guarantee that such currents will not appear in this altered equilibrium.

\section{Resonant radial electric field considerations}
\label{er_res}

In tokamaks, it is well known that rapid plasma rotation is possible in the toroidal direction as a result of symmetry. 
If one then assumes the ordering of $v_E\sim v_{ta}$, then radial electric fields are capable of producing sonic flows.
The radial electric field that corresponds to sonic rotation is known as the resonant electric field, which in axisymmetry is $E_r^{res}=r\iota Bv_{ta}/(Rc)$ \citep{Beidler08}.
We take the radial electric field here to be defined by $E_r\equiv-d\Phi/dr$.

Constraints on the symmetry of the magnetic field, however, prevent ordering the flow velocity with the thermal speed in generic stellarators \citep{Helander07}, as well as perfectly quasisymmetry ones \citep{Sugama11_rapid}.
The form of the drift kinetic equation that is solved in SFINCS uses the $v_E\sim\rho_{*a}v_{ta}$ ordering to avoid the symmetric restrictions to the magnetic field that result from sonic flows.
From the SFINCS ordering, the vast majority of parameter regimes, geometries, and species, yield ambipolar radial electric fields, $E_r^a$, that are considerably smaller than the resonant electric field magnitude.
However, the $m_a^{-1/2}$ dependence of the resonant electric field can cause the ordering to break down for heavy impurities under certain conditions.
Solving this issue completely would demand a reordering to derive a new form of the drift kinetic equation.
We do not attempt to tackle this problem here, but leave it to future work.

It is also very interesting and relevant to note that quasi-helically-symmetric (QH) configurations produce a considerably larger gap between $E_r^a$ and $E_r^{res}$ than quasi-axisymmetric (QA) configurations for otherwise identical plasma parameters.
The relative size of these electric field for a given simulation can be found by deriving an analogous expression for the resonant radial electric field in quasisymmetry. 

If we start by assuming the $v_E\sim v_{ta}$ ordering, then the $v_E$ and parallel streaming terms will be of the same order
\begin{equation}
    v_{\parallel}\bm{\hat{b}}\cdot\nabla\chi \sim \frac{c}{B^2}\bm{E}\times\bm{B}\cdot\nabla\chi,
\end{equation}
where $\chi=M\theta-N\zeta$.
The contravariant and covariant representations of the magnetic field are, respectively,
\begin{eqnarray}
    \bm{B} = \nabla\psi_t\times\nabla\theta + \iota\nabla\zeta\times\nabla\psi_t\\
    \bm{B} = L\nabla\psi_t + I\nabla\theta + G\nabla\zeta,
\end{eqnarray}
where $2\pi\psi_t$ is the toroidal flux, $L=L(\psi_t,\theta,\zeta)$ is some scalar, and as detailed in \citep{Helander14}
\begin{eqnarray}
    \int_{S_{\zeta}}\bm{J}\cdot\nabla\zeta\,\sqrt{g}\,\mathrm{d}\theta\,\mathrm{d}\psi_t = \frac{c}{2}I(\psi_t)\\
    \int_{S_{\theta}}\bm{J}\cdot\nabla\theta\,\sqrt{g}\,\mathrm{d}\zeta\,\mathrm{d}\psi_t = \frac{c}{2}G(\psi_t),
\end{eqnarray}
where $1/\sqrt{g}=(\nabla\psi_t\times\nabla\theta)\cdot\nabla\zeta$ is the Jacobian, and $S_{\zeta}$ and $S_{\theta}$ correspond to surfaces where $\zeta=const$ and $\theta=const$, respectively.
Solving for $E_r$ when $v_E\sim v_{\parallel}\sim v_{th}$ yields an expression for the resonant electric field
\begin{equation}
    E_r^{res} \sim \left|\frac{rv_{ta}B^2}{c}\frac{M\iota-N}{MG+NI}\right|.
\end{equation}
Typically, $I\ll G$, so the QH devices examined in the following sections (all of which have $|N|\ge 4$) allow one to run neoclassical codes at larger $E_r^a$ before the ordering breakdown is reached.
For this reason, only QH results are available in some SFINCS simulations with steep gradients and/or heavier impurity ions.
As a workaround for QA, we have only considered cases where $E_r^a/E_r^{res}<1/3$, in order to be sufficiently far from the resonance to avoid unreliable results due to the breakdown of $v_E\sim\rho_{*a}v_{ta}$.

\section{Magnetic field configurations}
\label{configs}

Throughout the remainder of this paper, we aim to provide results that are general to a wide range of quasisymmetric stellarators.
We have therefore chosen 8 distinct quasisymmetric stellarator configurations (summarized in table \ref{tab:tt}) to examine, some of which were designed to be QA, and the others QH.
Here, we have used the C09R00 equilibrium from NCSX, the Nuhrenberg configuration from figure 1 and table 1 in \citep{Nuhrenberg88}, and the quasi-helically-symmetric configuration of HSX.
HSX is the only configuration in this list that has been constructed to date.
To allow for a fair comparison between devices,
each device was scaled to the minor radius, $a$, and on-axis magnetic field, $B_0$, of the Henneberg et al QA configuration \citep{Quasdex}, $a=0.602$m and $B_0=2.10$T.

\begin{table}
\begin{center}
\def~{\hphantom{0}}
\begin{tabular}{lccc}
 \hline
 \multicolumn{4}{c}{Quasisymmetric Stellarators} \\
 \hline
 Configuration & QS     Type &   $N_{fp}$   & Aspect Ratio\\
 \hline
 Henneberg \citep{Quasdex} & QA & 2 & 3.40\\
 NCSX \citep{NCSX} & QA  & 3 & 4.37\\
 ARIES-CS \citep{ARIES} & QA & 3 & 4.56\\
 Wistell-Aten \citep{Wistell} & QH & 4 & 6.94\\
 HSX \citep{HSX} & QH & 4 & 10.17\\
 Garabedian \citep{Garabedian} & QA & 2 & 2.60\\
 Nuhrenberg-Zille \citep{Nuhrenberg88} & QH & 6 & 11.76\\
 CFQS \citep{CFQS1,CFQS2} & QA & 2 & 4.35\\
 \hline
\end{tabular}
\end{center}
\caption{Quasisymmetric stellarator configurations that have been studied in this work. QA-quasi-axisymmetric, QH-quasi-helically-symmetric, $N_{fp}$-Number of Field Periods.}
\label{tab:tt}
\end{table}

\section{Results}
\label{results}

The results generated below employ the full linearized Fokker-Planck collision operator of SFINCS, with two ion species.
The main ions are taken to be hydrogen in each of the simulations, while the charge and mass of the impurity ion can vary between runs.
It is assumed here that ion temperatures are equivalent, $T_i=T_z$, due to the fast equilibration time.
The choice of the temperature and density profiles in the following results is based on the modeling of an ECRH-heated, W7-X plasma in Fig 5 of \citep{Turkin11}.
The density gradient, however, is not determined from Fig 5 in \citep{Turkin11}, but rather chosen so as to give particular values of $\eta^{-1}$.
Further, the density gradient is taken to be equivalent between ion species $\nabla\ln n_i=\nabla\ln n_z$, or in other words, the profile of $Z_{eff}$ is flat.

The recent work of \citep{Helander17} has shown that temperature screening can be achieved in reactor-relevant, \textit{mixed-collisionality} plasmas (highly-collisional impurities and low-collisionality bulk ions) at large normalized radius $r_N=0.88$.
With the increased temperature in the core, however, it is possible that the bulk ions and impurities in reactor-grade plasmas will both have low collisionalities, depending of course on the particular impurity ion.
The picture for temperature screening becomes more pessimistic as collisionality decreases, which can be seen in Fig 1 and 2 in \citep{Helander17}.
For our purposes of understanding how much symmetry-breaking can be tolerated prior to losing this effect, we choose to study collisionalities below the region of temperature screening in \citep{Helander17}, in order to ensure that this transition will be observed in at least some cases.

\subsection{Impact of Magnetic Field Symmetry-Breaking on Impurity Particle Flux}
\label{eps_eff}
\subsubsection{Flat Density Profile: $\eta^{-1}=0$}

As one increases the magnitude of $\epsilon_{sb}$ from 0 up to the true magnetic field, the transport due to the helical wells will also increase. It is not clear a priori exactly how this incremental breaking of symmetry will change the impurity particle flux.
However, it is clear that for many cases of interest there should be some critical value, $\epsilon_{sb}^c$, where the radial impurity particle flux, $\Gamma_z$, changes sign, which will depend on the particular magnetic equilibrium.

In this section, we examine the $\epsilon_{sb}$ dependence of $\Gamma_z$ for each of the configurations in table \ref{tab:tt} in select parameter regimes.
It should be understood that the magnitude of $\Gamma_z$ is less important than the sign in this section.

\begin{figure}
\centering
\includegraphics[width=5in]{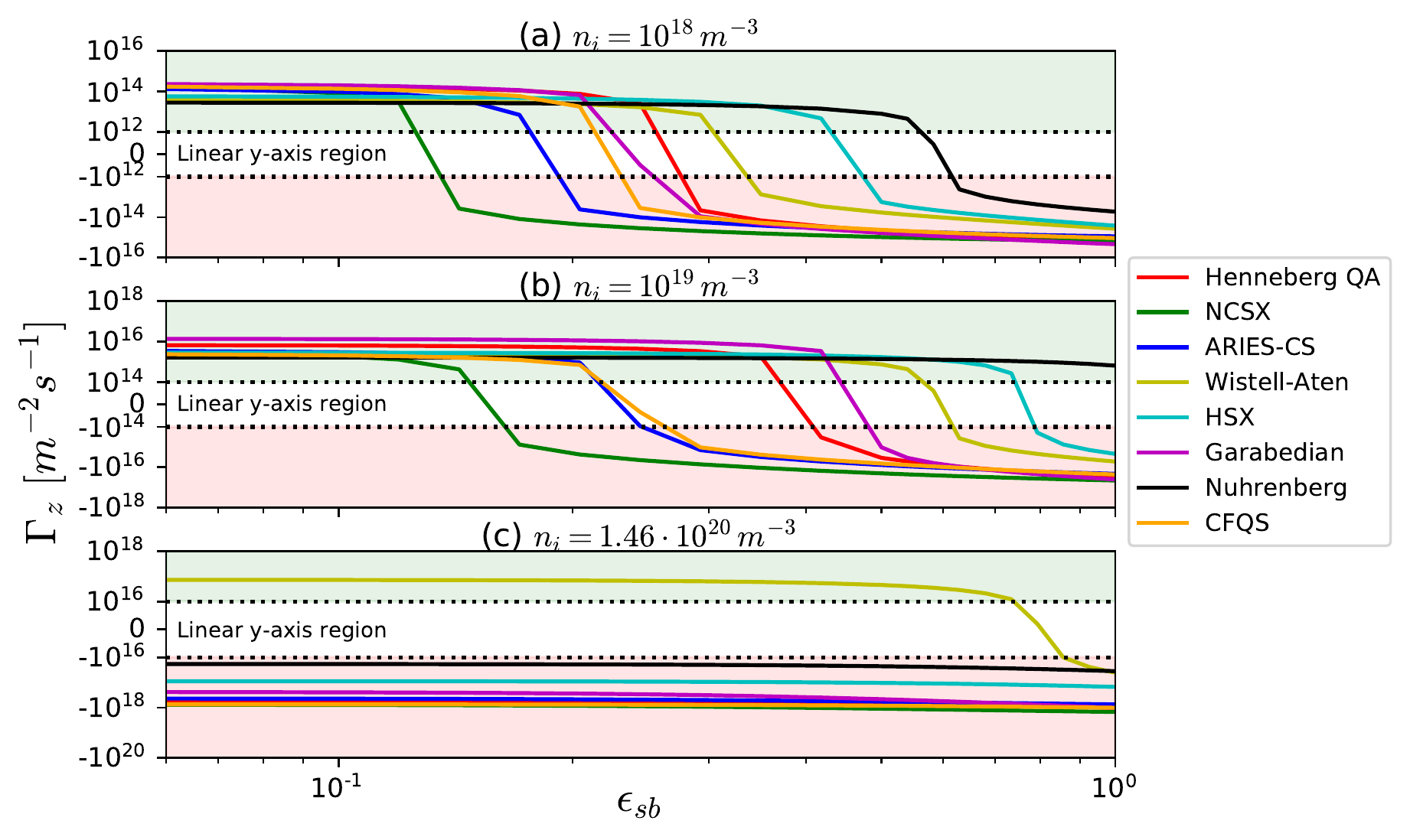}
\caption{(Color online) The impurity particle flux at $\eta^{-1}=0$ for $C^{6+}$ is plotted as a function of the symmetry-breaking amplitude at a normalized radius of $r_N=0.25$. $T=4\,$keV, $dT/dr=-0.97\,$keV/m, and $Z_{eff}=2$ were kept constant for all subplots.
The upper, green-shaded region denotes positive $\Gamma_z$ (impurity screening).
The lower, red-shaded region corresponds to negative $\Gamma_z$ (impurity accumulation). The collisionalities for each subplot are (a) $\nu_*^{z}=2.26\cdot 10^{-4}R/a$, (b) $\nu_*^{z}=2.26\cdot 10^{-3}R/a$, and (c) $\nu_*^{z}=3.29\cdot 10^{-2}R/a$.}
\label{fig:epssb_25}
\end{figure}

The $E_r$ for each simulation was chosen to be the ambipolar $E_r$ for the $\epsilon_{sb}=1$ case, considering that $E_r$ becomes progressively less important in calculating radial fluxes as the magnetic field approaches symmetry.
It also becomes difficult to accurately calculate the radial electric field for small values of $\epsilon_{sb}$.
We choose fully-ionized carbon, $C^{6+}$, as the impurity in Figures \ref{fig:epssb_25} and \ref{fig:epssb_50} in order to avoid proximity to the resonant electric field in all configurations (see Section \ref{er_res}).
Finally, we take $\alpha=\sum_{a\ne i}n_aZ_a^2/(n_iZ_i^2)=1$, corresponding to $Z_{eff}=2$.

In Figure \ref{fig:epssb_25}(a)-(c), we have plotted results at $r_N\simeq0.25$ for all devices at increasing values of collisionality, which is achieved by varying the ion density at constant temperature.
We define the normalized ion and impurity collisionalities by $\nu_*^{i}\equiv \nu_{i}R/v_{ti}$ and $\nu_*^z\equiv\nu_{z}R/v_{tz}$, respectively.
Here, and in the results that follow, $r_N\equiv r/a = \sqrt{\psi_t/\psi_a}$, where $a$ is the minor radius, and $2\pi\psi_a$ is the toroidal flux at the last closed flux surface (computed in VMEC \citep{VMEC}).
At this radial location we take $T_i=T_z=4\,$keV and $dT_i/dr=dT_z/dr=-0.97\,$keV/m.
In the lowest-collisionality case of Figure \ref{fig:epssb_25}(a) (with $n_i=10^{18}\,\textrm{m}^{-3}$), there is a similar $\epsilon_{sb}$ dependence of $\Gamma_z$ for each of the devices, regardless of the type of quasisymmetry (QA or QH).
For a magnetic field with near-perfect quasisymmetry ($\epsilon_{sb}=10^{-2}$), the resulting $\Gamma_z$ is positive, indicating a presence of the temperature screening effect.
As $\epsilon_{sb}$ is increased, $\Gamma_z$ decreases until eventually changing sign at some value of $\epsilon_{sb}<1$.

The first thing that can be understood from this plot is that at this collisionality, none of the devices that were studied displayed temperature screening at the actual magnetic field, $\epsilon_{sb}=1$.
However, the value of $\epsilon_{sb}^c$ where temperature screening is lost will depend on the magnetic configuration.
In the case of Nuhrenberg-Zille, for example, the transition occurs at $\epsilon_{sb}^c\simeq 0.6$, which is essentially saying that the symmetry-breaking terms must be $\sim60\%$ of their actual values to ensure temperature screening under these conditions. 
Toward the left side of the plot, the NCSX transition occurs at $\epsilon_{sb}^c\simeq 0.1$, requiring the symmetry-breaking terms to be $\sim$10x smaller. 

\begin{figure}
\centering
\includegraphics[width=3in, trim={0.5in 2.0in 0.5in 3.0in}]{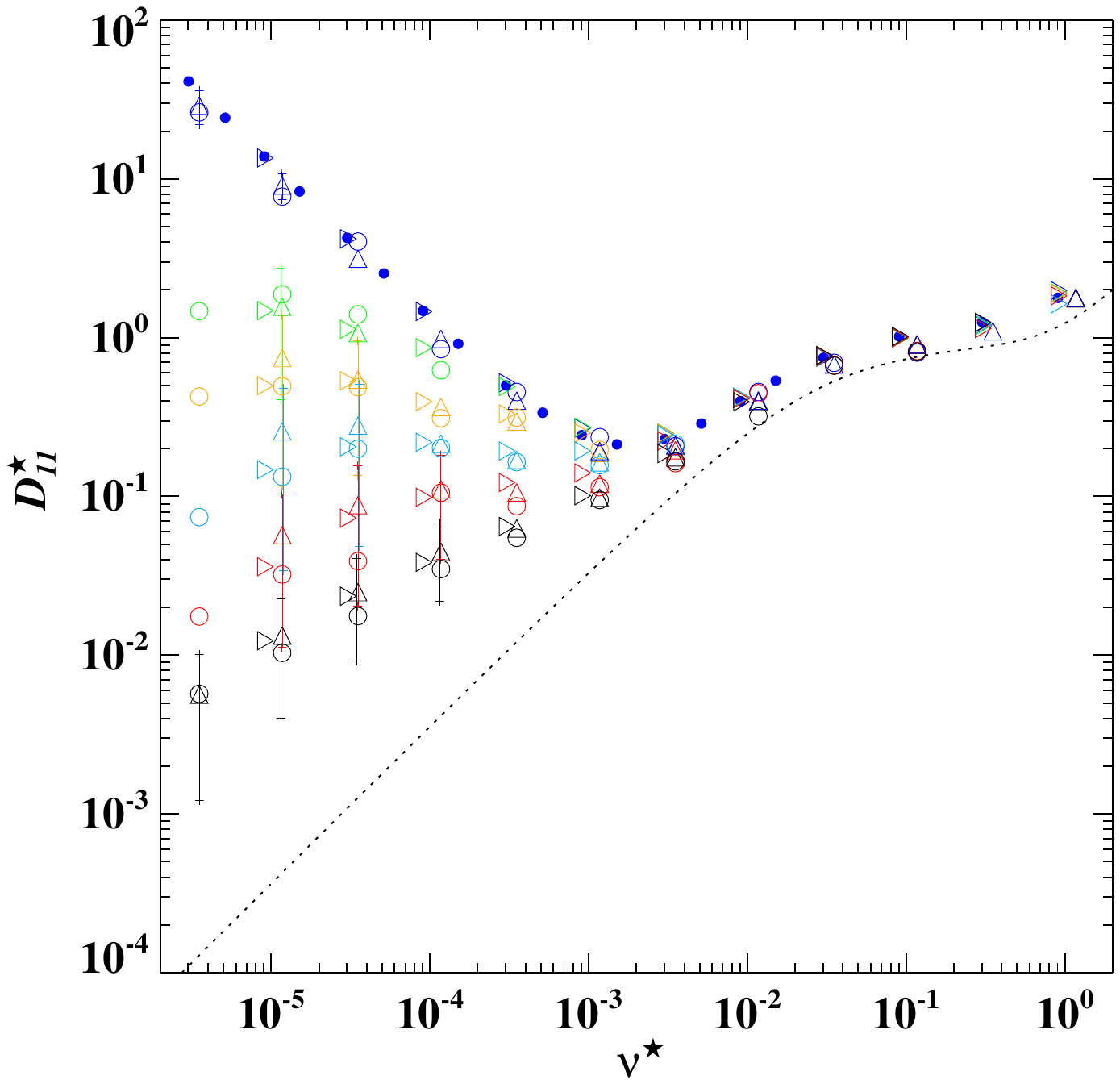}
\caption{(Color online) Definitions of normalizations, markers/colors, and details can be found in Figure (14) of \citep{Beidler11}. The $D_{11}$ transport coefficient is plotted as a function of collisionality for HSX geometry.
Here, the colors represent different $v_E$ values.
DKES \citep{DKES_1,DKES_2} results are depicted by  triangles $(\triangle)$, NEO-2 \citep{Nemov99,Kernbichler08} by filled circles $(\bullet)$, and Monte-Carlo results are plotted using open circles $(\bigcirc)$ \citep{Tribaldos01}, and right-point triangles $(\triangleright)$ \citep{Allmaier08}.
The dotted line is a simulation with equivalent perfect helical symmetry and $E_r=0$.}
\label{fig:beidler}
\end{figure}

\begin{figure}
\centering
\includegraphics[width=5in]{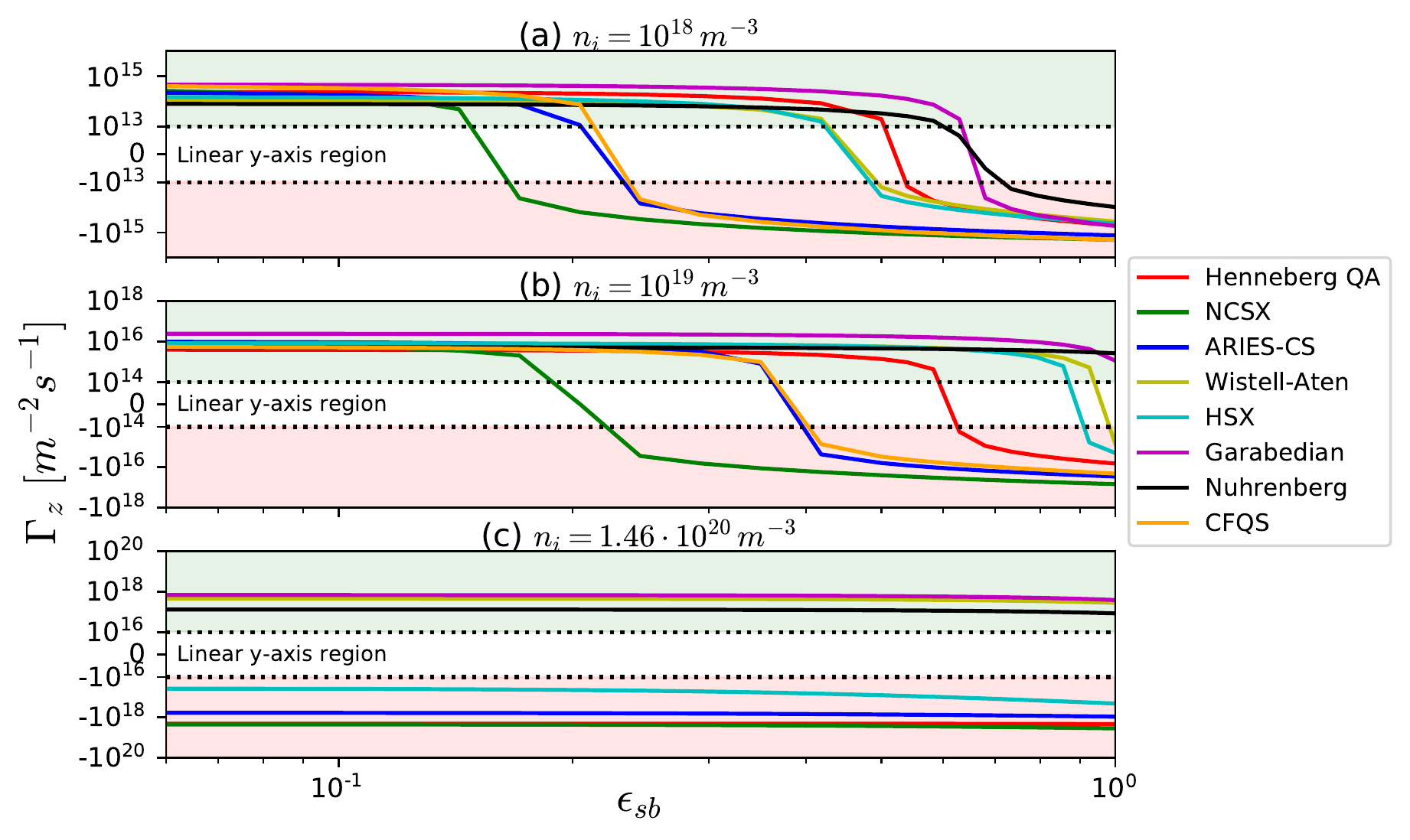}
\caption{(Color online) The impurity particle flux at $\eta^{-1}=0$ for $C^{6+}$ is plotted as a function of the symmetry-breaking amplitude at a normalized radius of $r_N=0.50$. $T=3.3\,$keV, $dT/dr=-4.78\,$keV/m, and $Z_{eff}=2$ were kept constant for all subplots. 
The upper, green-shaded region denotes positive $\Gamma_z$ (impurity screening).
The lower, red-shaded region corresponds to negative $\Gamma_z$ (impurity accumulation).
The collisionalities for each subplot are (a) $\nu_*^{z}=3.32\cdot 10^{-4}R/a$, (b) $\nu_*^{z}=3.32\cdot 10^{-3}R/a$, and (c) $\nu_*^{z}=5.32\cdot 10^{-2}R/a$.}
\label{fig:epssb_50}
\end{figure}

In Figure \ref{fig:epssb_25}(b), the same plot as in Figure \ref{fig:epssb_25}(a) is constructed, however, the density (and hence collisionality) has been increased by an order of magnitude.
First, it should be remarked that at this collisionality, the Nuhrenberg-Zille configuration actually achieves temperature screening at $\epsilon_{sb}=1$.
While this is the only such configuration to do so, it is also true that $\epsilon_{sb}^c$ has increased for each configuration from the respective values in Figure \ref{fig:epssb_25}(a).
Since $\epsilon_{sb}^c$ can approximate closeness to quasisymmetry, it follows that increasing the collisionality appears to improve the ``effective quasisymmetry" of a flux surface.
The meaning behind the term ``effective quasisymmetry" can be understood from a figure in \citep{Beidler11}, which we have reused in Figure \ref{fig:beidler} with permissions.
At low collisionality, there is a difference (depending on $E_r$) between the $D_{11}$ coefficient (describing radial transport) for HSX and the perfectly quasisymmetric case, indicating a sensitivity of the particles to the exact structure of the magnetic field.
As collisionality is increased, this difference becomes less pronounced as the contribution to transport from helically trapped particles decreases.
At a high-enough collisionality, $D_{11}$ in Figure \ref{fig:beidler} is about the same for perfect quasisymmetry as it is for HSX, regardless of $E_r$.
Noting that the magnetic trapping structures can be quite different in perfect symmetry and a nominally quasisymmetric field, the similarities in $D_{11}$ indicate a decreased sensitivity of the particles to the exact structure of the magnetic field at higher collisionalities.
Thus, in the context of quasisymmetry, increasing the collisionality brings the transport closer to symmetric levels, ``effectively" increasing quasisymmetry.

However, there is a limit to the beneficial impacts of increasing the density, as can be seen in Figure \ref{fig:epssb_25}(c), where $n_i=1.46\cdot10^{20}\,\textrm{m}^{-3}$.
Aside from the Wistell-Aten configuration, all of the other configurations at near-perfect quasisymmetry do not display an outward impurity flux.
There are two possible explanations for why this might take place in perfect symmetry.
First, for QA configurations, it is possible that $E_r^a$ and $E_r^{res}$ are close enough that ambipolarity no longer holds, and the higher-order $E_r$ terms become important.
To explain this effect in QH, one must recall that temperature screening in axisymmetry is not predicted at high collisionalities \citep{Rutherford74}, except for cases where $\alpha\rightarrow 0$ for collisional ions \textit{and} impurities.
In Appendix \ref{app}, it can be seen that beyond some critical $\nu_*^{ii}$ in axisymmetry, the impurity flux becomes negative for most $\eta^{-1}$.
Figure \ref{fig:epssb_25}(c) is thus indicating that we are hovering around that critical collisionality where temperature screening is not possible, even in perfect symmetry.

The situation is less pessimistic in Figure \ref{fig:epssb_50}, where we look at $r_N=0.50$ with $T=3.3\,$keV and $dT/dr=-4.78\,$keV/m.
The trends are largely similar to Figure \ref{fig:epssb_25} at each collisionality, however, there are a handful of cases where temperature screening can be seen at $\epsilon_{sb}=1$.
Furthermore, at the highest collisionality, the Wistell-Aten, Garabedian, and Nuhrenberg configurations have an outward impurity particle flux for all $\epsilon_{sb}$.
The collisionality at $r_N=0.50$ is only slightly higher than at $r_N=0.25$, and $\eta^{-1}=0$ at both $r_N$, so the differences at these radii are likely caused by the distinct magnetic field modes, $B_{mn}$.

\begin{figure}
\centering
\includegraphics[width=3in]{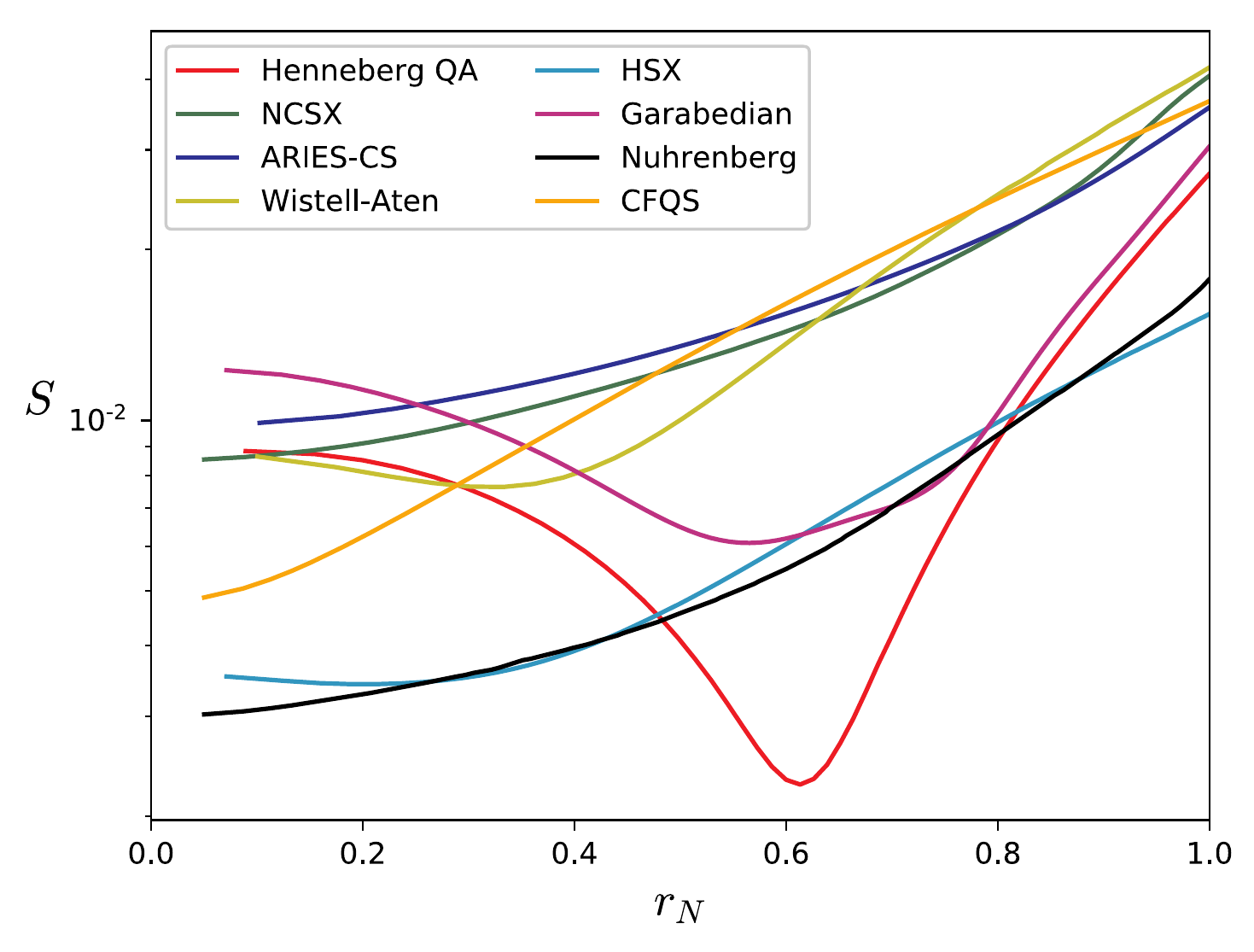}
\caption{(Color online) The amplitude of the symmetry-breaking terms, $S$, are plotted as a function of normalized radius, $r_N$.}
\label{fig:S_rN}
\end{figure}

The differences between Figures \ref{fig:epssb_25} and \ref{fig:epssb_50} can be understood by looking at the magnitude of symmetry-breaking terms
\begin{equation}
    S\equiv\sqrt{\sum_{m,n\ne mN}B_{mn}^2/B_{00}^2},
    \label{eq:s}
\end{equation}
as a function of $r_N$ in Figure \ref{fig:S_rN}.
If we consider the curves for Henneberg QA and Garabedian, we can compare the difference in the values of $\epsilon_{sb}^c$ in Figure \ref{fig:epssb_25}(a) and Figure \ref{fig:epssb_50}(a).
In moving from $r_N=0.25$ to $r_N=0.50$ in Figure \ref{fig:S_rN}, the symmetry-breaking amplitude for Henneberg QA and Garabedian decreases by $\sim 4$.
The corresponding increase in $\epsilon_{sb}^c$ from $r_N=0.25$ to $r_N=0.50$ is $\sim 2-3$ for both configurations.
If we were to then consider the respective CFQS curves, there is an increase in $S$ between these radii of $\sim 2$, where a \textit{decrease} in $\epsilon_{sb}^c$ is observed.
This presents a connection between the closeness to quasisymmetry of a flux surface, and the realization of temperature screening.
The remaining configurations have a difference in $S$ of less than a factor of 2 at these radii, and $S$ is also larger at $r_N=0.50$.
Unlike the connection between $S$ and the change in $\epsilon_{sb}^c$ for Henneberg QA, Garabedian, and CFQS,  the change in $\epsilon_{sb}$ is \textit{positive} (although small) for the remaining configurations.
This could be potentially be accounted for by a complicated dependency on collisionality, $E_r$, and the aspect ratio.

The overall results in Figures \ref{fig:epssb_25} and \ref{fig:epssb_50} indicate that for reactor-relevant plasma parameters, temperature screening could be achievable in certain configurations.
However, as it is unlikely that the density profile will be completely flat, it is imperative to understand how $\Gamma_z$ varies with $\eta^{-1}$.

\subsubsection{Finite Peaked Density Gradients: $\eta^{-1}>0$}
\label{finite_eta}

Peaking of the main ion's density profile drives an inward neoclassical impurity flux. This result is shown for axisymmetry or quasisymmetry in Appendix \ref{app}. In non-symmetric stellarators, the situation is complicated by not only the presence of the radial electric field as a driving gradient in the impurity flux, but the fact that $L_{11}$ depends on $E_r$, and in different ways depending on the collisionality regime.
An exact analytic solution for $\Gamma_z$ is therefore intractable in most cases.
However, it is possible to approximate the solution by using a similar procedure to that used in \citep{Velasco17}, but generalizing for arbitrary $Z_{eff}$.
We start with an expression for the particle flux of an arbitrary species that is valid far from quasisymmetry at low collisionality
\begin{equation}
    \Gamma_a=-n_aL_{11}^a\left[\frac{1}{n_a}\frac{dn_a}{dr}-\frac{Z_aeE_r}{T_a}+\delta_a\frac{1}{T_a}\frac{dT_a}{dr}\right].
    \label{eq:gamma_all}
\end{equation}
It is possible to then explicitly solve the ambipolarity condition $\sum_aZ_a\Gamma_a=0$ for $E_r$.
If we take $T_i=T_z=T_e$, it is possible to drop the contribution of $\Gamma_e$ to the ambipolar condition since $L_{11}^e\ll L_{11}^i$.
Finally, if the density $(n_a'/n_a)$ and temperature $(T_a'/T_a)$ gradients are taken to be equivalent for the bulk ions and impurities, one can solve for the radial electric field
\begin{equation}
    \frac{eE_r}{T} = \frac{\left(L_{11}^i+\frac{\alpha}{Z}L_{11}^z\right)\frac{n'}{n}+\left(L_{11}^i\delta_i+\frac{\alpha\delta_z}{Z}L_{11}^z\right)\frac{T'}{T}}{L_{11}^i+\alpha L_{11}^z}.
\end{equation}
By plugging this back into the expression for $\Gamma_z$
\begin{equation}
    \Gamma_z = n_z L_{11}^z\left[\frac{L_{11}^i(Z-1)}{L_{11}^i+\alpha L_{11}^z}\frac{n'}{n}+A\frac{T'}{T}\right],
    \label{eq:gamma_eta}
\end{equation}
where $A$ is a scalar, and \citep{Beidler11} has shown that $L_{11}^a>0$ in all cases.
In the approximation that $L_{11}^a$ and $A$ do not vary strongly with $n'/n$ (incorrect for accurately computing $\Gamma_a$), Eq \ref{eq:gamma_eta} indicates that for $n'/n<0$, the density gradient will always have an unfavorable effect on impurity accumulation, and one that worsens as $|n'/n|$ increases.

\begin{figure}
\centering
\includegraphics[width=3in]{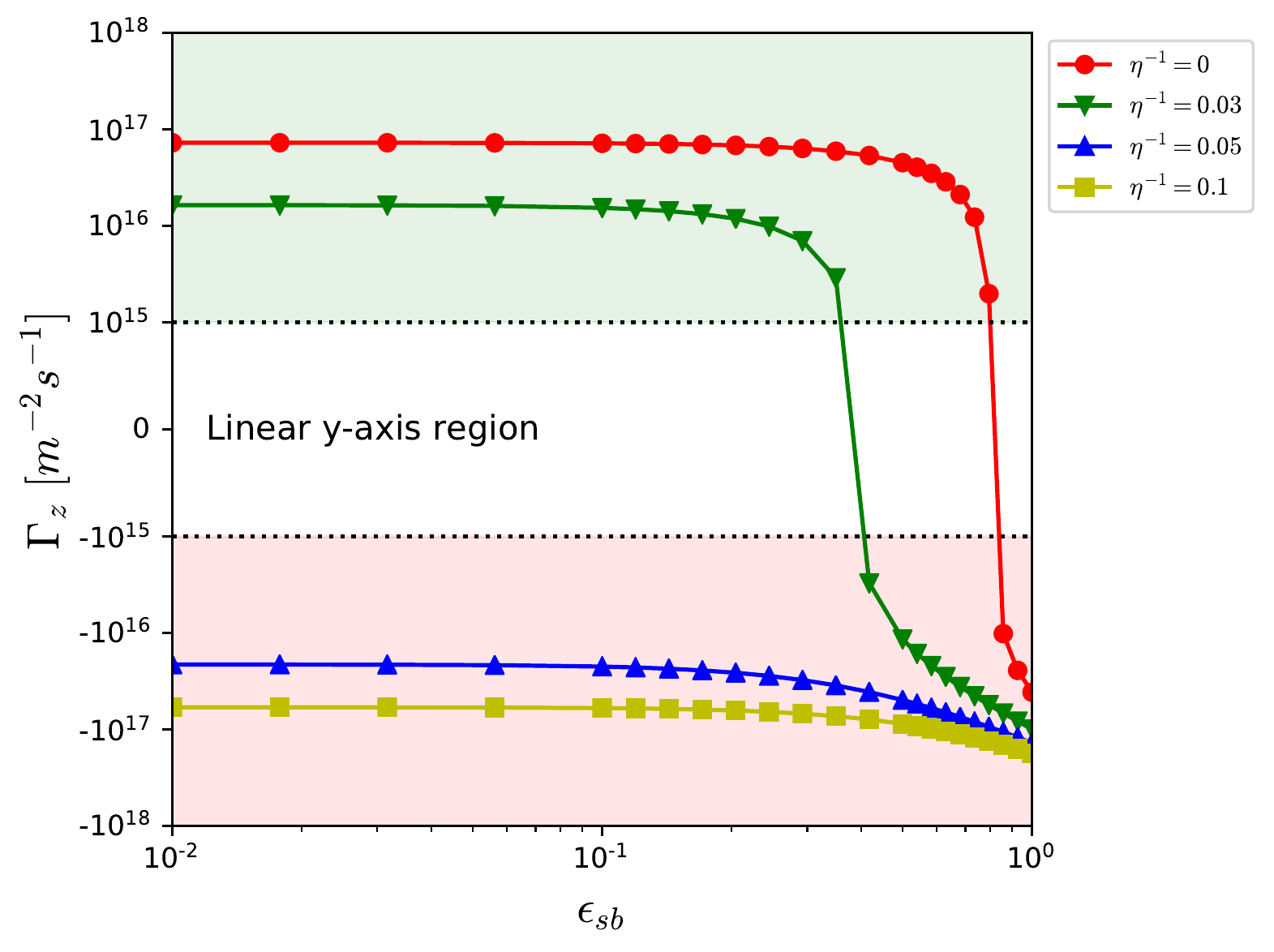}
\caption{(Color online) The impurity particle flux for $C^{6+}$ in Wistell-Aten is plotted as a function of $\epsilon_{sb}$, the scaling factor for the symmetry-breaking terms, at $r_N=0.25$. Each curve represents a different relative density gradient, but the physical parameters are otherwise identical to those of Figure \ref{fig:epssb_25}(c).
The upper, green-shaded region denotes positive $\Gamma_z$ (impurity screening).
The lower, red-shaded region corresponds to negative $\Gamma_z$ (impurity accumulation).}
\label{fig:wistell_eta}
\end{figure}

This is evident in Figure \ref{fig:wistell_eta} in the context of how $\Gamma_z$ is affected by $\epsilon_{sb}$.
Each curve in the figure was calculated with the Wistell-Aten configuration at various $\eta^{-1}$.
The red curve is the $\eta^{-1}=0$ case (identical to the Wistell curve in Figure \ref{fig:epssb_25}(c)), where the degree of quasisymmetry is nearly good enough to retain temperature screening at such parameters.
If a small density gradient $\eta^{-1}=0.03$ is introduced, $\epsilon_{sb}^c$ decreases by nearly a factor of 2.
Any further increase in $\eta^{-1}$ pushes the plasma to the point where even perfect quasisymmetry cannot support temperature screening.
This makes the situation of temperature screening even more pessimistic, because even if $\Gamma_z>0$ in a particular collisionality regime, simply introducing a density gradient can flip the sign.

\begin{figure}
\centering
\begin{subfigure}[t]{0.5\textwidth}
\includegraphics[width=2.5in, trim={1.5in 3.0in 1.5in 3.0in}]{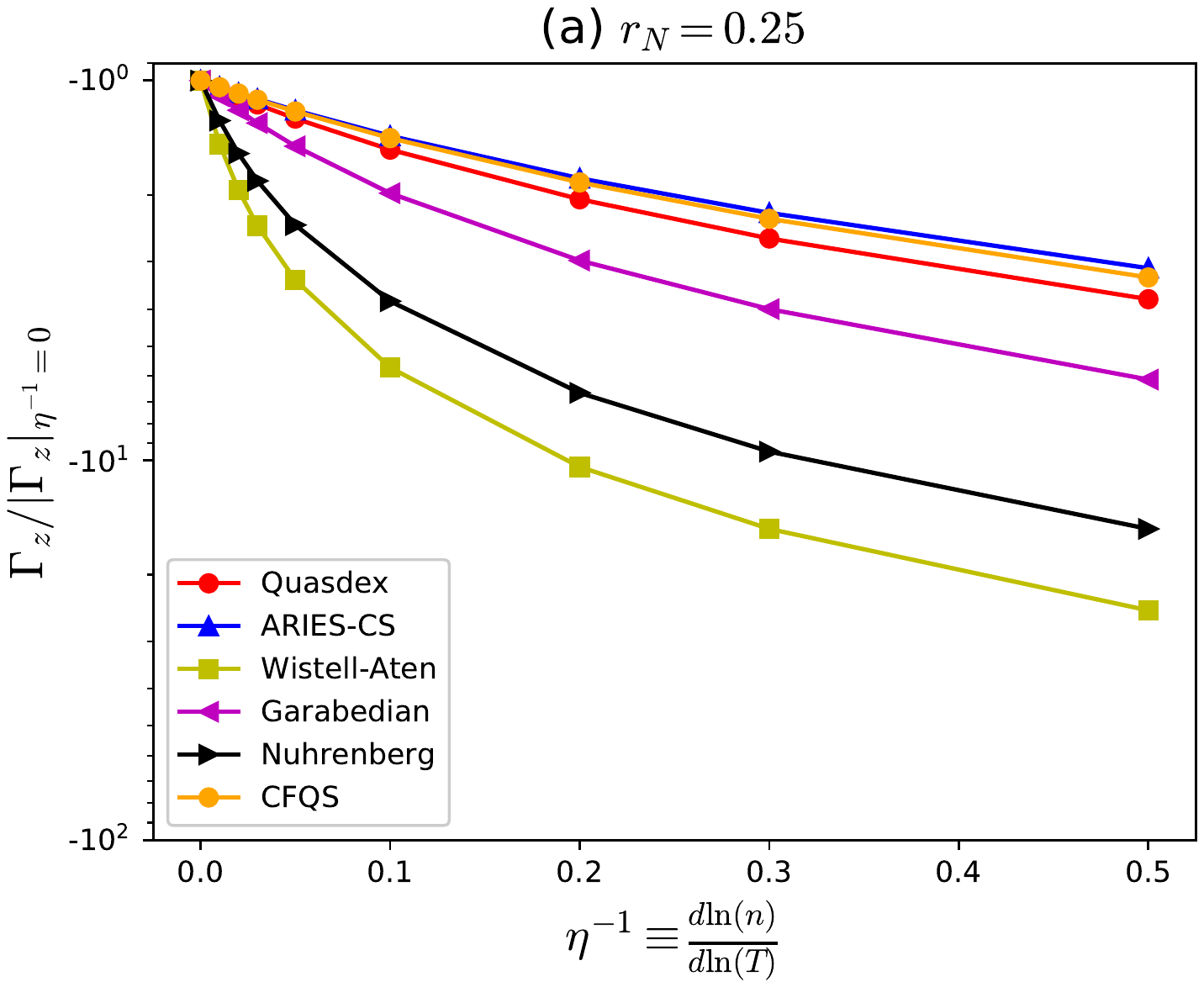}
\end{subfigure}
\begin{subfigure}[t]{0.45\textwidth}
\includegraphics[width=2.5in, trim={1.5in 3.0in 1.5in 3.0in}]{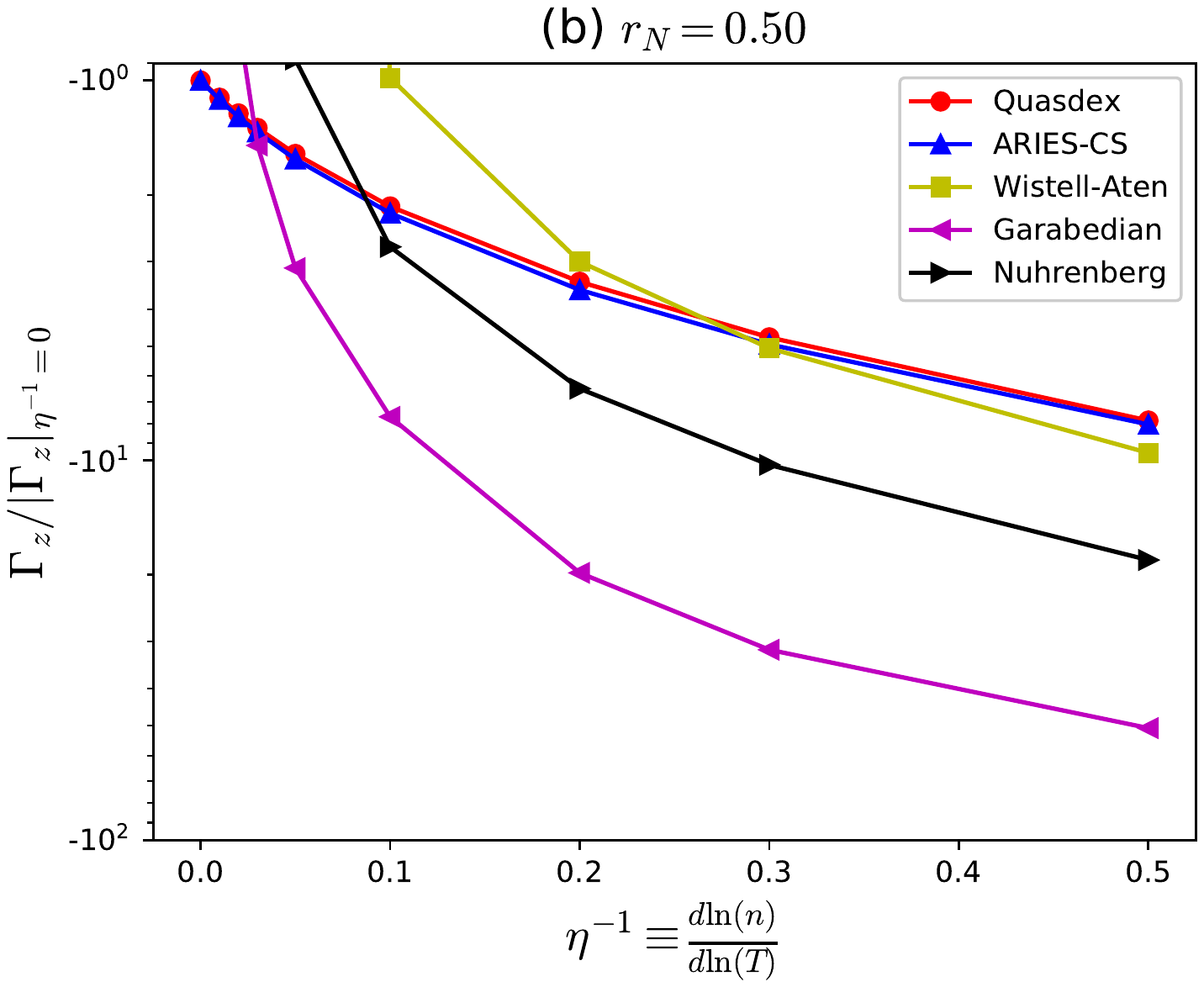}
\end{subfigure}
\caption{(Color online) The impurity particle flux for $C^{6+}$ has been normalized to its magnitude at $\eta^{-1}=0$ and plotted as a function of $\eta^{-1}$ at (a) $r_N=0.25$, and (b) $r_N=0.50$. Every simulation was performed at the true magnetic field $\epsilon_{sb}=1$, with the $E_r^a$ independently calculated at every $\eta^{-1}$. The physical parameters are otherwise identical to those of Figures \ref{fig:epssb_25}(c) and \ref{fig:epssb_50}(c).
The data points above $-10^0$ in (b) are those corresponding to devices with positive $\Gamma_z$ at $\epsilon_{sb}=1$, thus giving a value of $+10^0$ at $\eta^{-1}=0$.}
\label{fig:finite_eta}
\end{figure}

In all likelihood, there will be some finite density gradient in a reactor-relevant plasma, likely corresponding to an inward flux of impurities.
It is then of interest to see how the magnitude of $\Gamma_z$ changes, relative to its value at $\eta^{-1}=0$, as the strength of the density gradient is increased.
In Figure \ref{fig:finite_eta}, the ratio $\Gamma_z/|\Gamma_z|_{\eta^{-1}=0}$ is plotted as a function of $\eta^{-1}$ for various configurations, where each simulation was calculated at the true magnetic field, and used its own $E_r^a$.
In every case shown in Figure \ref{fig:finite_eta}(a), $\Gamma_z$ is negative and a decreasing function of $\eta^{-1}$, indicating that increasing the strength of the peaked density gradient will intensify impurity accumulation.
In a scenario where the length scale of the density gradient is only twice that of the temperature gradient $(\eta^{-1}=0.5)$, the enhancement in $\Gamma_z$ at this radius can be increased by a factor of $\sim20$.
The picture appears to at least slightly worsen at $r_N=0.50$ in Figure \ref{fig:finite_eta}(b), where the enhanced accumulation has close to doubled from the values in Figure \ref{fig:finite_eta}(a) in most cases.

\subsection{Comparison to Turbulent Fluxes}
\label{turbulence}

In this section, we use results from the parameter scans in the previous section to compare the neoclassical particle flux, $\Gamma_z$, and heat flux, $Q_{total}=Q_i+Q_z$, to a gyro-Bohm estimate for turbulent transport, $\Gamma_z^{turb}\sim n_z D_{gb}|\nabla T|/T$, and $Q_{total}^{turb}\sim D_{gb}|\nabla T|(n_i+n_z)$.
In these expressions, $D_{gb}=\rho_*^2v_{ti}a$ is the gyro-Bohm diffusion coefficient, where we have taken the minor radius to be the relevant length scale.

\begin{figure}
\centering
\begin{subfigure}[t]{0.50\textwidth}
\includegraphics[width=2.7in]{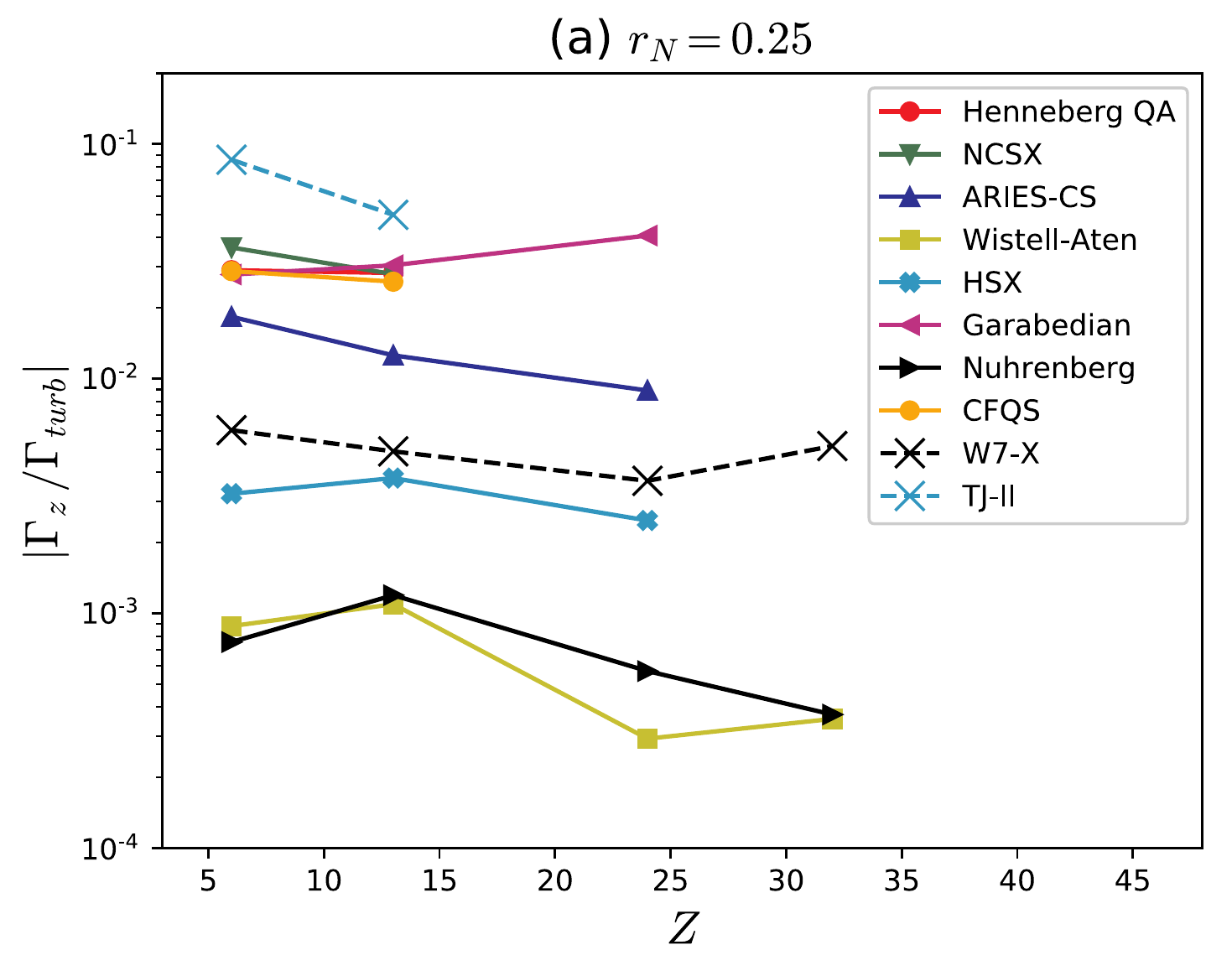}
\end{subfigure}
\centering
\begin{subfigure}[t]{0.45\textwidth}
\includegraphics[width=2.7in]{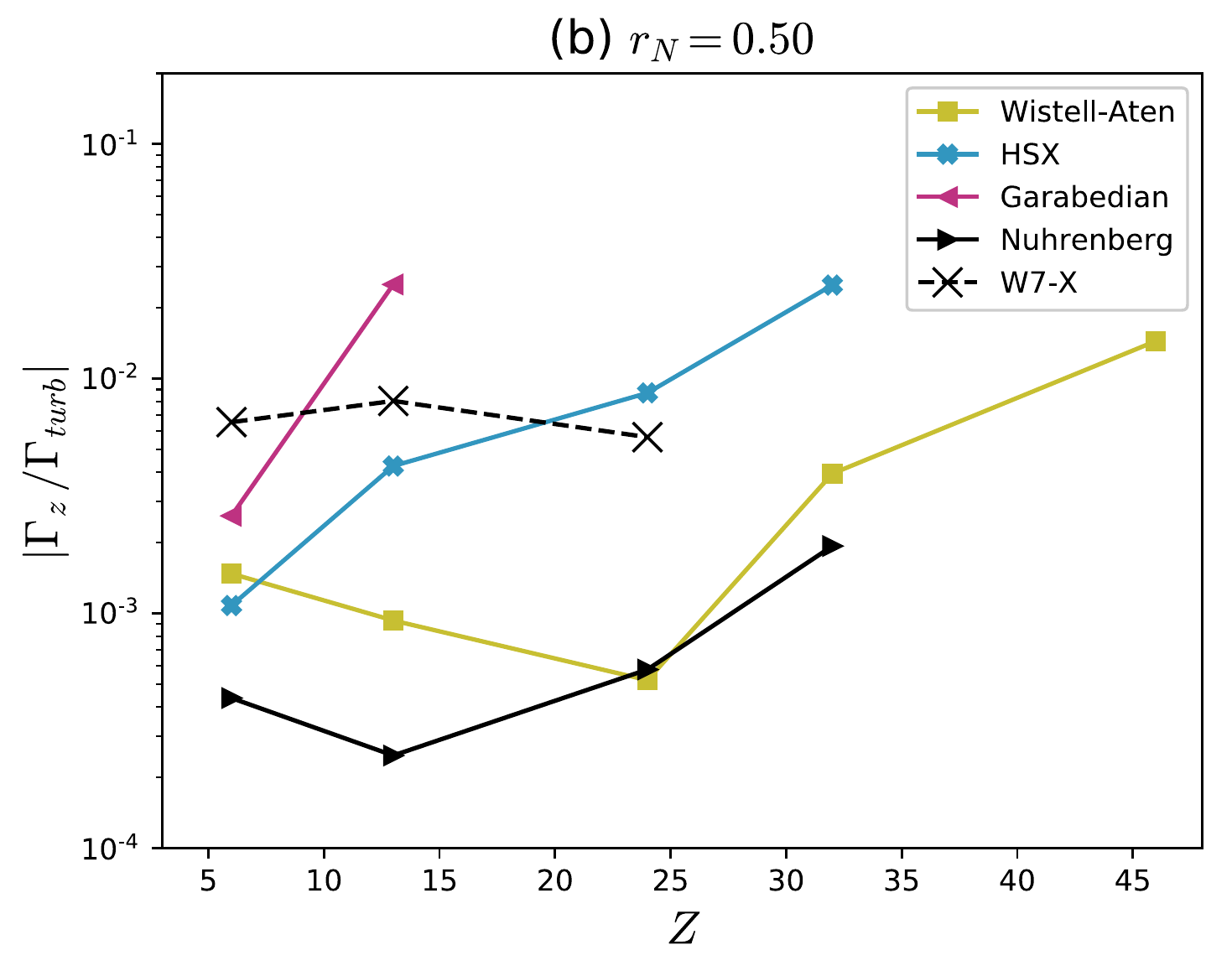}
\end{subfigure}
\caption{(Color online) The neoclassical impurity particle flux at $\eta^{-1}=0$ has been normalized to a gyro-Bohm estimate of the turbulent impurity particle flux (see text). This ratio is plotted as a function of the impurity charge (and mass) for (a) $r_N=0.25$, and (b) $r_N=0.50$. Plasma parameters correspond to those of Figures \ref{fig:epssb_25}(c) and \ref{fig:epssb_50}(c).}
\label{fig:gamma_turb}
\end{figure}

Figure \ref{fig:gamma_turb} examines how the neoclassical particle fluxes compare to $\Gamma_z^{turb}$ as a function of the impurity ion charge for each device, using the equilibrium magnetic field, $\epsilon_{sb}=1$.
Only flat density profiles $(\eta^{-1}=0)$ are considered in Figure \ref{fig:gamma_turb}.
In Figure \ref{fig:gamma_turb}(a), we look at the impurity particle flux for $r_N=0.25$, where the temperature gradients are weaker.
At this radial location, the ratio of fluxes does not have a consistent strong trend with $Z$.
This is observed in both QA and QH configurations, as well as the \textit{non}-quasisymmetric TJ-II \citep{TJII} and W7-X devices, which have been included since the results of Figure \ref{fig:epssb_25} do not scale the amplitude of symmetry-breaking modes.
The most striking feature of Figure \ref{fig:gamma_turb}(a), however, is the dominance of turbulent transport.
Of the quasisymmetric configurations that were studied, the \textit{largest} calculated neoclassical flux at $r_N=0.25$ is only $\sim 5\%$ of the turbulent value.
These small ratios indicate that regardless of whether temperature screening is present at a given collisionality, the turbulence will probably control the sign of the particle flux.

At $r_N=0.50$ in Figure \ref{fig:gamma_turb}(b), the overall sensitivity of this ratio to the impurity species in QA is unclear since the larger gradients push $E_r^a$ close enough to $E_r^{res}$ that results are unreliable (see Section \ref{er_res}).
Only configurations with at least 2 points have been shown in Figure \ref{fig:gamma_turb}, eliminating all but 1 QA configuration.
Apart from W7-X, there is an eventual point for each configuration at which further increase in $Z$ corresponds to an increase in the relative importance of neoclassical fluxes.
Even with this increase in the ratio, the neoclassical contribution to the radial particle flux is $<3\%$ of the turbulent value.

It should be reiterated here that these results have been generated with a flat density profile.
While it is still unknown exactly how the density profiles will behave in a reactor, it is likely that $|\eta^{-1}| > 0$.
From Figure \ref{fig:finite_eta}, it can then be inferred how this neoclassical to turbulence ratio will change if a peaked density gradient is introduced.
In a non-ideal scenario, where $\eta^{-1}=0.5$, the ratio could increase by more than a factor of 10, depending on the configuration.
At $r_N=0.25$, this would still only lead to the neoclassical flux being $\sim 10\%$ of the turbulence for most configurations.

\begin{figure}
\centering
\includegraphics[width=3.5in]{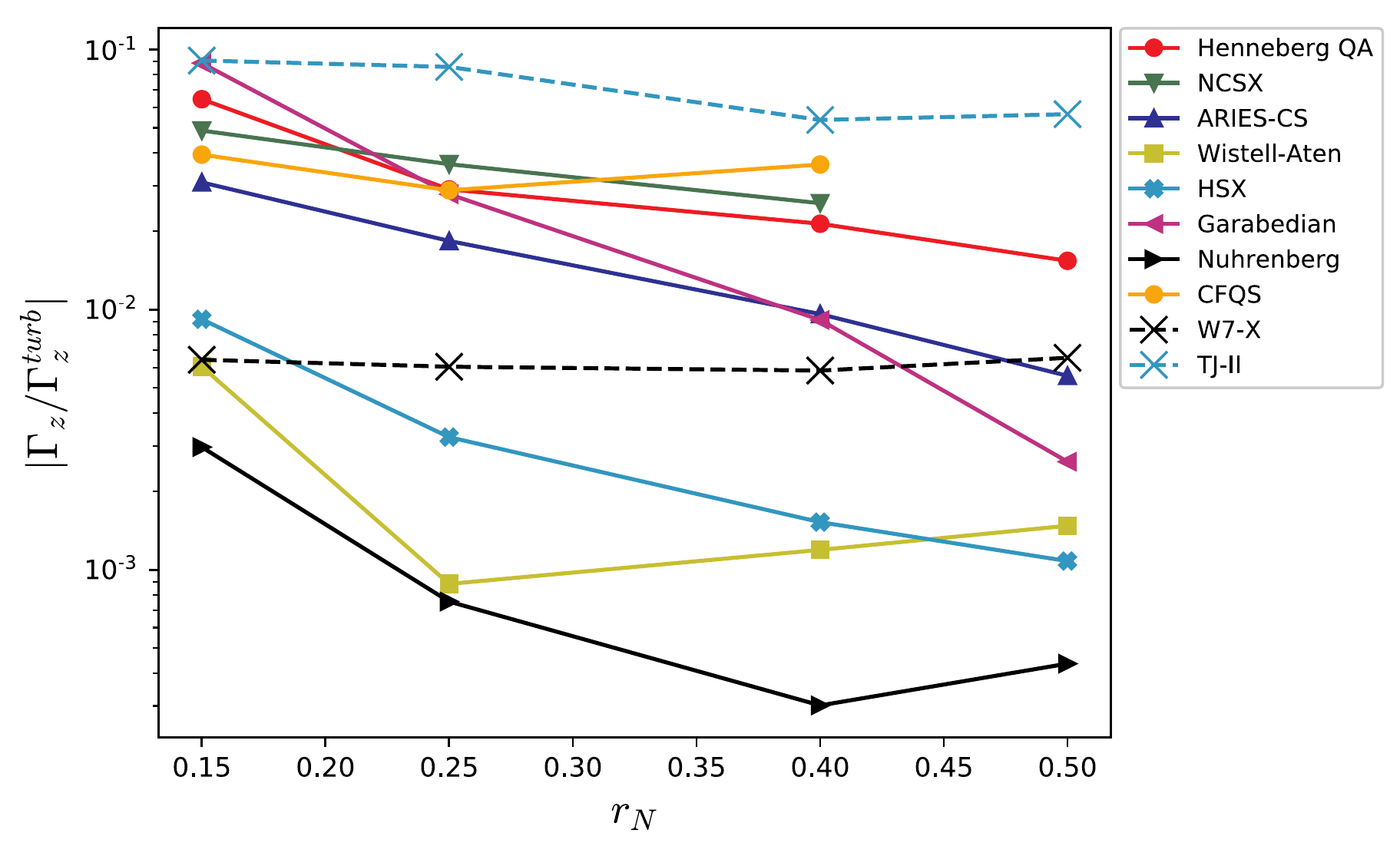}
\caption{(Color online) The neoclassical impurity particle flux at $\eta^{-1}=0$ for $C^{6+}$ has been normalized to a gyro-Bohm estimate of the turbulent impurity particle flux (see text). This ratio is plotted as a function of the normalized radius $r_N$. Plasma profiles at $r_N=0.25$ and $r_N=0.50$ correspond to those of Figures \ref{fig:epssb_25}(c) and \ref{fig:epssb_50}(c), respectively. At $r_N=0.15$: $T=4.1\,$keV, $dT/dr=-0.58\,$keV/m, $n_i=1.51\cdot 10^{20}\,\textrm{m}^{-3}$. At $r_N=0.40$: $T=3.75\,$keV, $dT/dr=-3.88\,$keV/m, $n_i=1.43\cdot 10^{20}\,\textrm{m}^{-3}$.}
\label{fig:gamma_radial}
\end{figure}

It is also of practical importance to understand how this ratio of neoclassical to turbulent particle flux varies with distance from the magnetic axis.
This radial profile is shown in Figure \ref{fig:gamma_radial}, where the radial points $r_N=0.15$ and $r_N=0.40$ (profiles can be found in the caption of Figure \ref{fig:gamma_radial}) have been added to the previously calculated values at $r_N=0.25$ and $r_N=0.50$.
For most but not all, the ratio tends to either decrease or remain constant as one moves out radially, indicating that turbulence becomes increasingly more important.
This follows experimental observations \citep{Canik07,Pablant18} that show neoclassical fluxes at negligible levels when compared to turbulence far from the magnetic axis.

\subsubsection{Total Heat Flux}

Along with particle fluxes, it is also of great practical importance to compare the neoclassical and turbulent \textit{heat fluxes} at different locations within the plasma.
If the dominant transport channel can be identified, this can better inform future efforts in optimizing for a certain type of transport over a particular radial domain.
In this section we do not distinguish between ion and impurity heat fluxes, since we primarily care about the \textit{total} heat flux (ion+impurity) that is crossing a flux surface.

\begin{figure}
\centering
\includegraphics[width=3.5in]{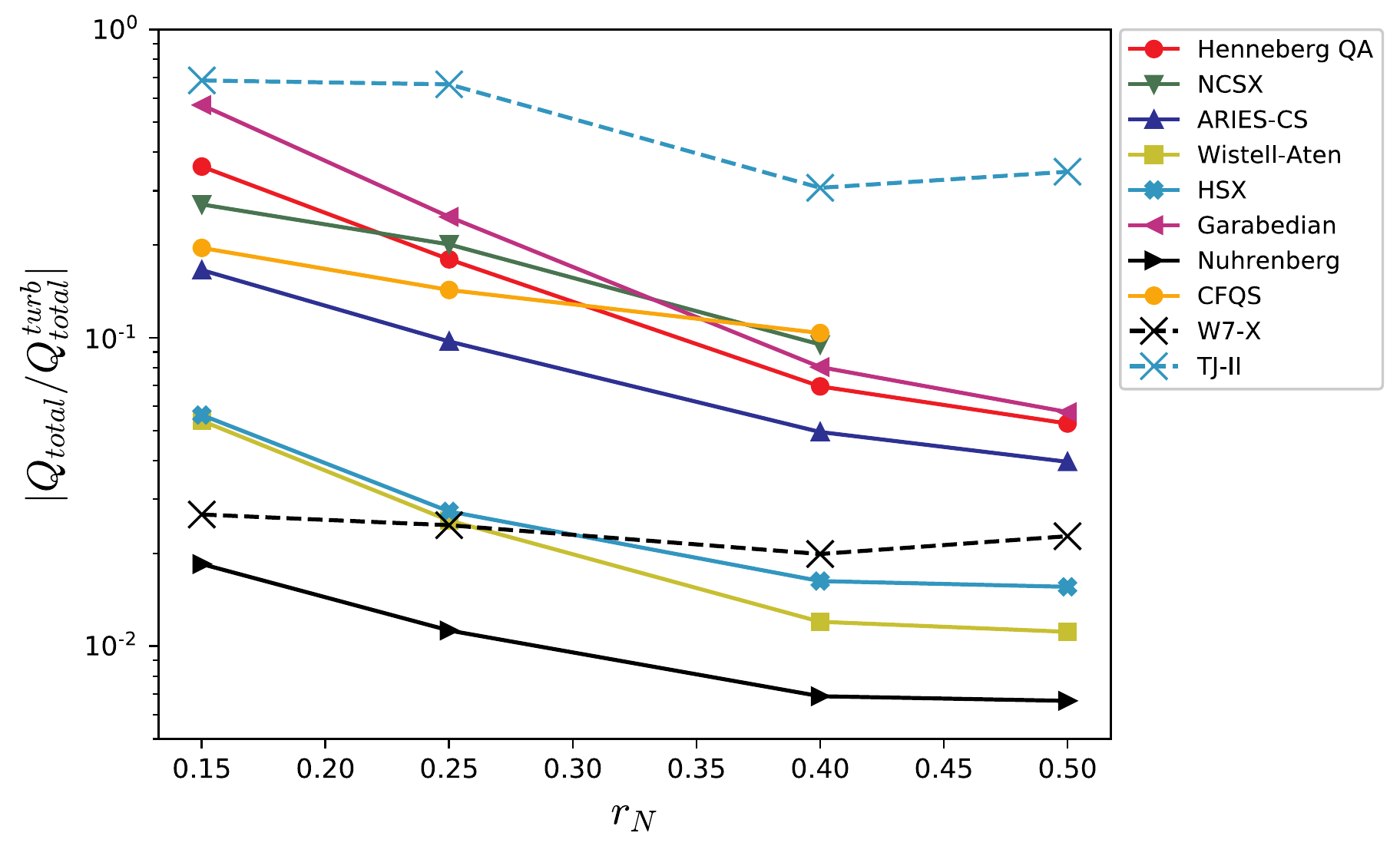}
\caption{(Color online) The total (ion+impurity) heat flux at $\eta^{-1}=0$ for $C^{6+}$ has been normalized to a gyro-Bohm estimate of the total turbulent heat flux (see text). This ratio is plotted as a function of the normalized radius $r_N$. Plasma profiles are the same as for Figure \ref{fig:gamma_radial}.}
\label{fig:hflux_radial}
\end{figure}

Thus, shifting our attention to the ratio of neoclassical to turbulent heat fluxes, the results in Figure \ref{fig:hflux_radial} show the radial profiles of this ratio for each configuration.
The overall trend is similar to Figure \ref{fig:gamma_radial}, except that the magnitude of this ratio is a bit higher than the respective points in Figure \ref{fig:gamma_radial}, especially closer to the magnetic axis.

However, it is important to mention here that unlike the impurity particle flux, we have found this ratio to be independent of $\eta^{-1}$, and the particular impurity species.
So while the ratios in Figure \ref{fig:gamma_radial} may appear smaller in comparison, a heavy impurity in the presence of a density gradient could change that. 
This is to say that these heat flux ratios are more robust over a wider range of potential reactor-relevant parameters than the impurity particle flux.

The general trend of the decreasing relative importance of neoclassical heat flux compared to turbulence with respect to radius is in agreement with experimental results \citep{Pablant18,Canik07}. 
With that said, for $r_N\geq 0.25$, the neoclassical heat flux is, at best, 30\% of the turbulent value, and in many cases this ratio is even smaller.

These magnitudes appear to be at odds with Figure 7 in \citep{Pablant18}, where neoclassical simulations (SFINCS) of an ECRH-heated W7-X experiment show that the neoclassical electron heat flux constitutes $\sim 65\%$ of the input power through the flux surface at $r_N=0.25$.
If the remaining flux is presumed to be turbulence-driven, then the neoclassical electron heat flux should be about twice the turbulent value.
By comparing this neoclassical result to a gyro-Bohm estimate using $\rho_i=2.04\cdot 10^{-3}\,$m, and $L_{T_e}\equiv (1/T_e|dT_e/dr|)^{-1} = 0.66\,$m in the expression $Q_e^{turb}\sim n_e\rho_{*i}^2v_{ti}a(T_eL_{T_e}^{-1})$ one finds $|Q_e/Q_e^{turb}|\sim 0.1$.
This disagreement underlines the nature of gyro-Bohm as only an \textit{estimate} of turbulence.
Setting the coefficient of $D_{gb}$ to $1$ for every configuration and set of plasma parameters is bound to yield results that can differ by an appreciable amount relative to the actual turbulent fluxes.

\section{Effective Helical Ripple as a Quasisymmetry Metric}
\label{ripple}

\begin{figure}
\centering
\begin{subfigure}[t]{0.5\textwidth}
\includegraphics[width=2.5in, trim={1.5in 3.0in 1.5in 3.0in}]{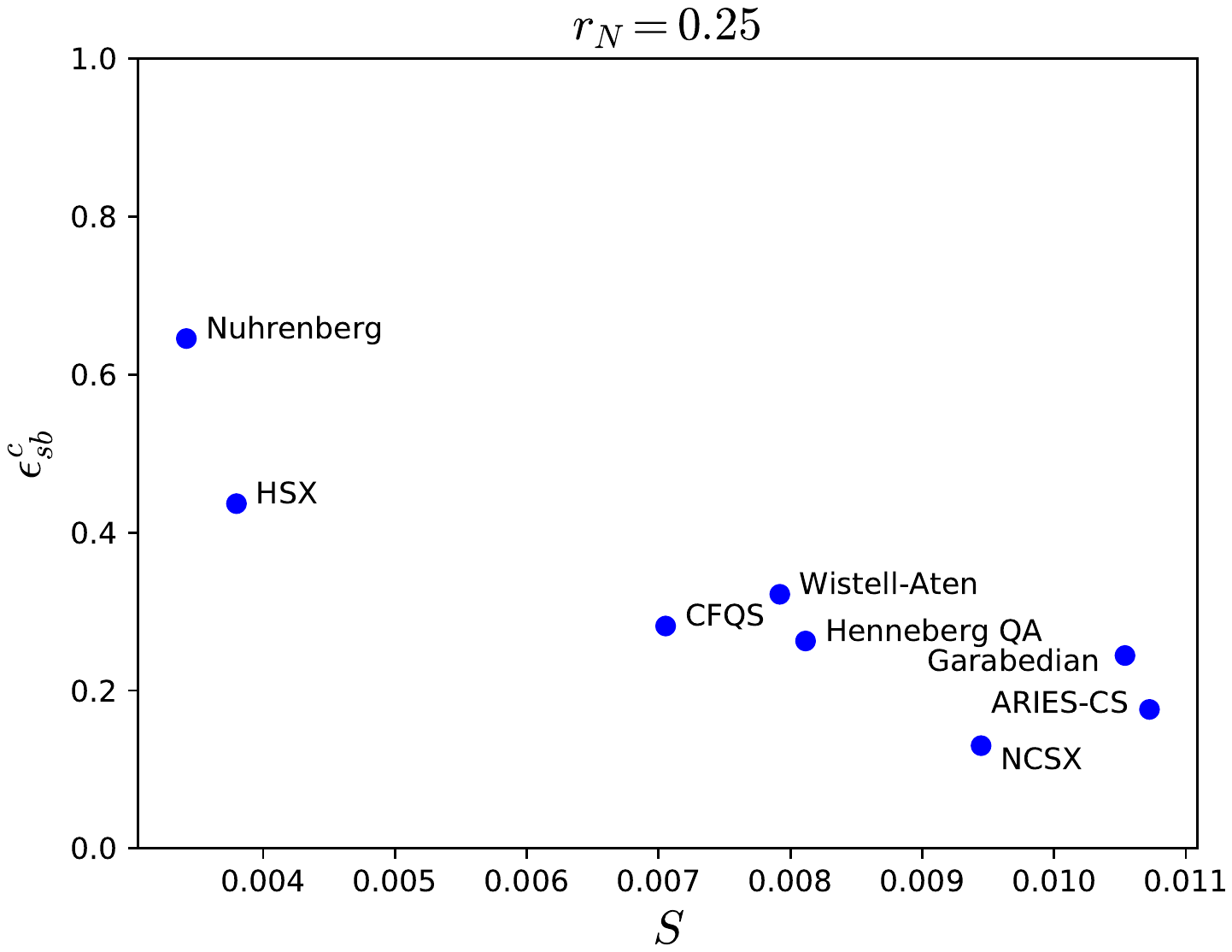}
\end{subfigure}
\begin{subfigure}[t]{0.45\textwidth}
\includegraphics[width=2.5in, trim={1.5in 3.0in 1.5in 3.0in}]{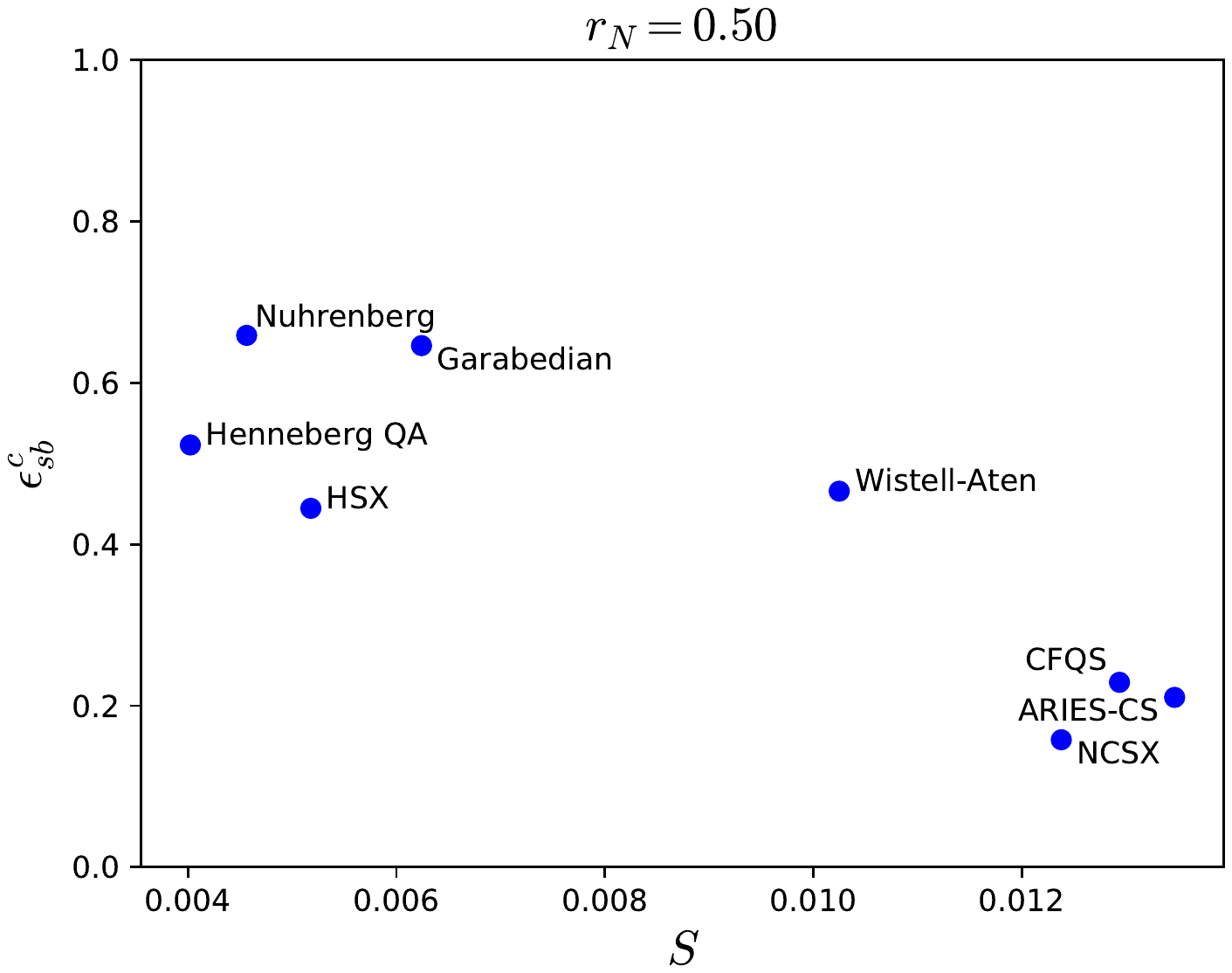}
\end{subfigure}
\caption{The critical symmetry-breaking parameter $\epsilon_{sb}^c$ for each configuration as a function of the corresponding $S$ value has been plotted at (a) $r_N=0.25$, and (b) $r_N=0.50$, which correspond to the $\epsilon_{sb}^c$ values from Figure \ref{fig:epssb_25}(a) and Figure \ref{fig:epssb_50}(a), respectively.}
\label{fig:epssb_S}
\end{figure}

From Section \ref{eps_eff}, we showed how there was a connection between $S$ and $\epsilon_{sb}^c$ that helped to explain how $\epsilon_{sb}^c$ changed between the two flux surfaces that were studied.
This connection can be seen more clearly in Figures \ref{fig:epssb_S}(a)-(b), where the value of $\epsilon_{sb}^c$ from Figures \ref{fig:epssb_25}(a) and \ref{fig:epssb_50}(a) has been plotted as a function of $S$ on the respective surfaces for each of the configurations.
For both $r_N=0.25$ and $r_N=0.50$, there is a visible anti-correlation between the two quantities even when considering that these configurations have very different properties.
It thus seems reasonable to expect that minimizing $S$ on a flux surface will increase $\epsilon_{sb}^c$.

Along with $S$, the effective helical ripple, $\epsilon_{eff}$, is sometimes taken to be a metric for quasisymmetry that could be used for stellarator optimization.
$\epsilon_{eff}$, which is a measure of neoclassical transport in the $1/\nu$ regime, was computed with the NEO code \citep{Nemov99}.

\begin{figure}
\centering
\includegraphics[width=2.7in, trim={1.5in 3.0in 1.5in 3.0in}]{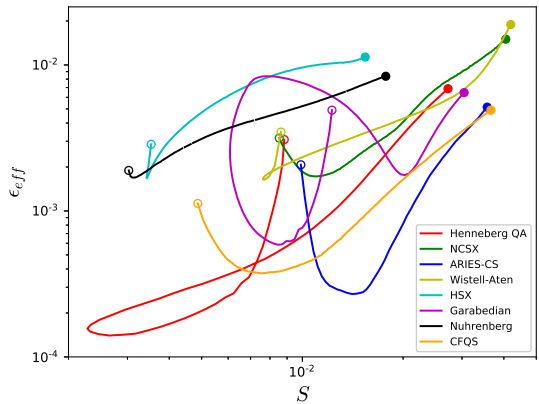}
\caption{(Color online) The effective helical ripple (calculated with NEO \citep{Nemov99}) is plotted as a function of the amplitude of the symmetry-breaking terms. The open circles here denote the value on-axis ($r_N=0$). The closed circles correspond to the value at $r_N=1$. These curves do not change with plasma parameters.}
\label{fig:epseff_S}
\end{figure}

In Figure \ref{fig:epseff_S}, the effective helical ripple is plotted as a function of $S$ for each configuration.
A number of these curves are multi-valued, indicating a non-monotonic change in quasisymmetry from the magnetic axis to the last closed flux surface (LCFS).
To clarify the radial dependency of each curve, the open circle at the end of a curve denotes the magnetic axis, $r_N=0$, and a closed circle the LCFS, $r_N=1$.
It can be seen \citep{Quasdex,Mynick06} that if individual symmetry-breaking $B_{mn}$ harmonics are plotted as a function of $r_N$, that the amplitude tends to increase with distance from the magnetic axis.
Indeed, this trend can be seen for a handful of configurations in Figure \ref{fig:epseff_S}, indicating a correlation between $\epsilon_{eff}$ and the closeness to quasisymmetry.
However, this is decidedly not universal among QA configurations.
Henneberg QA, for example, has a symmetry-breaking amplitude that decreases by nearly an order of magnitude from $r_N=0\rightarrow r_N\simeq0.6$, and then increases again to a value at $r_N=1$ that is larger than its value at the $r_N=0$. 

Moreover, this monotonicity in $S$, or lack thereof, is not necessarily tied to the value of $\epsilon_{eff}$.
Returning to Henneberg QA as an example, the initial decrease in $S$ with $r_N$ is accompanied with a decrease in $\epsilon_{eff}$.
Then, the subsequent increase in $S$ corresponds to an increase in $\epsilon_{eff}$, indicating a possible correlation between $S$ and $\epsilon_{eff}$.
However, the behavior is different in the core of ARIES-CS, where the smallest value of $S$ corresponds to a relatively large value of $\epsilon_{eff}$, which decreases considerably as $S$ is \textit{increased}.
The point here is that while there are certainly configurations where $\epsilon_{eff}$ scales with $S$, it is just as likely that they may not correlate well at all, and the assumption that small $\epsilon_{eff}$ indicates good quasisymmetry cannot be justified a priori.
It has in fact been shown in \citep{Cary97} that one can achieve omnigeneity $(\epsilon_{eff}=0)$ far from quasisymmetry.

It is further interesting to note that in the cases where $\epsilon_{eff}$ does \textit{not} scale with $S$, the radial location where this disagreement happens is usually within $r_N\simeq 0.5$.
Above this $r_N$ (or in some configurations, a position much closer to the magnetic axis), the scaling of $\epsilon_{eff}$ with $S$ can be observed in every case.
An interpretation of this behavior is left to future work.

\section{Conclusions}

In this work, we have examined how impurity particle flux and the temperature screening effect are influenced by varying the closeness of the magnetic field to perfect quasisymmetry.
For realistic departures from symmetry, at very low values of collisionality and a flat density gradient, temperature screening was not observed for any quasisymmetric configuration.
However, with increasing collisionality one can see an increase in the ``effective quasisymmetry" of a flux surface.
This can lead to temperature screening in some cases.
Unfortunately, there is an upper limit to the benefits of higher collisionalities which, if surpassed, will lead to impurity accumulation even in perfect quasisymmetry.

When peaked density gradients are introduced, there is an overall negative effect on the impurity particle flux.
Increasing the density gradient peaking $(\eta^{-1}>0)$ enhances the strength of the impurity accumulation, and also leads to a reduction in the ``effective quasisymmetry".
Overall, while temperature screening is technically possible at the true magnetic field in select cases, it is unlikely to be present over the majority of reactor-relevant regimes.

The magnitudes of these results at the true magnetic field ($\epsilon_{sb}=1$) were then compared to a gyro-Bohm estimate for the turbulent fluxes.
Even in the non-ideal scenario of $\eta^{-1}=0.5$, the majority of configurations show neoclassical impurity particle fluxes that don't exceed 10\% of the respective turbulent flux, even for highly charged impurities.
While one should not discount possible changes when $\Phi_1$ is included, this presents a silver lining to the above statement predicting impurity accumulation in most scenarios.
If, as our results suggest, the turbulence ultimately determines the sign of the impurity particle flux in most experimental scenarios, the presence of temperature screening (a neoclassical phenomenon) would become mostly irrelevant to determining the direction of impurity particle transport.

Finally, it was shown that the value of $\epsilon_{eff}$ is not necessarily correlated with the amplitude of the symmetry-breaking terms on a flux surface.
Therefore, while $\epsilon_{eff}$ is a useful metric for the optimization of \textit{transport} in stellarator design, it is not the correct metric for optimizing quasisymmetry on a flux surface.

The authors would like to thank S Henneberg for providing the configuration in \citep{Quasdex}, A Bader for providing Wistell-Aten, J Schmitt for providing HSX, N Pomphrey for providing NCSX and ARIES-CS, and S Okamura for providing CFQS.
We would also like to thank Ian Abel, Alessandro Geraldini, Rogerio Jorge, and Elizabeth Paul for useful comments and discussions.
This work was supported by the U.S. Department of Energy, Office of Science, Office of Fusion Energy Science, under award DE-FG02-93ER54197.
This research used resources of the National Energy Research Scientific Computing Center (NERSC), a U.S. Department of Energy Office of Science User Facility operated under Contract No. DE-AC02-05CH11231.

\appendix

\section{Impurity Temperature Screening in Axisymmetry and Quasisymmetry}
\label{app}

A number of papers \citep{Connor73,Rutherford74,Rosenbluth72,Galeev68} have examined neoclassical transport quantities in the axisymmetric limit and can provide a more intuitive look at the properties of a plasma that can affect temperature screening.
These results are also applicable in perfect quasisymmetry.
In \citep{Connor73}, an expression for the bulk ion (taken to be hydrogen here) particle flux in the presence of electrons and a single impurity was derived, using a momentum-conserving, pitch-angle scattering collision operator, where all species are taken to be in the banana regime

\begin{equation}
    \Gamma_i = - C_i\left\{ \frac{T_i}{T_e}\left[\frac{n_i'}{n_i}-\left(\frac{0.09+0.5\alpha}{0.53+\alpha}\right)\frac{T_i'}{T_i}\right] - \frac{T_z}{ZT_e}\left[\frac{n_z'}{n_z}-0.17\frac{T_z'}{T_z}\right]\right\}.
\end{equation}
Here, $C_i$ is a positive coefficient that is independent of thermodynamic gradients, $Z$ is the impurity charge, and $\alpha =n_zZ^2/n_i = Z_{eff}-1$ represents the effect of the impurities on the transport coefficients.
Typically, one can take $T_i=T_z=T$, and we further assume the profiles of each species to be equal: $T_i'/T_i=T_z'/T_z=T'/T$, and $n_i'/n_i=n_z'/n_z=n'/n$.
An expression for the \textit{impurity} particle flux can then easily be obtained by satisfying the ambipolarity condition, $\sum_aZ_a\Gamma_a=0$.
Based on $\Gamma_e\sim\sqrt{m_e/m_i}\,\Gamma_i$, if we take $\Gamma_e\simeq 0$, then $\Gamma_z\simeq-\Gamma_i/Z$.
The impurity particle flux can now be shown to be

\begin{equation}
    \Gamma_z=\frac{C_i}{Z}\left\{\frac{T}{ZT_e}\left[L_a\frac{n'}{n}-L_b\frac{T'}{T}\right]\right\},
    \label{eq:gamma_axi}
\end{equation}
where $L_a\equiv Z-1$ and $L_b\equiv\{[Z(0.09+0.5\alpha)/(0.53+\alpha)]-0.17\}$. $L_b$ is always positive for $Z>1$.

\begin{figure}
\centering
\includegraphics[width=3in]{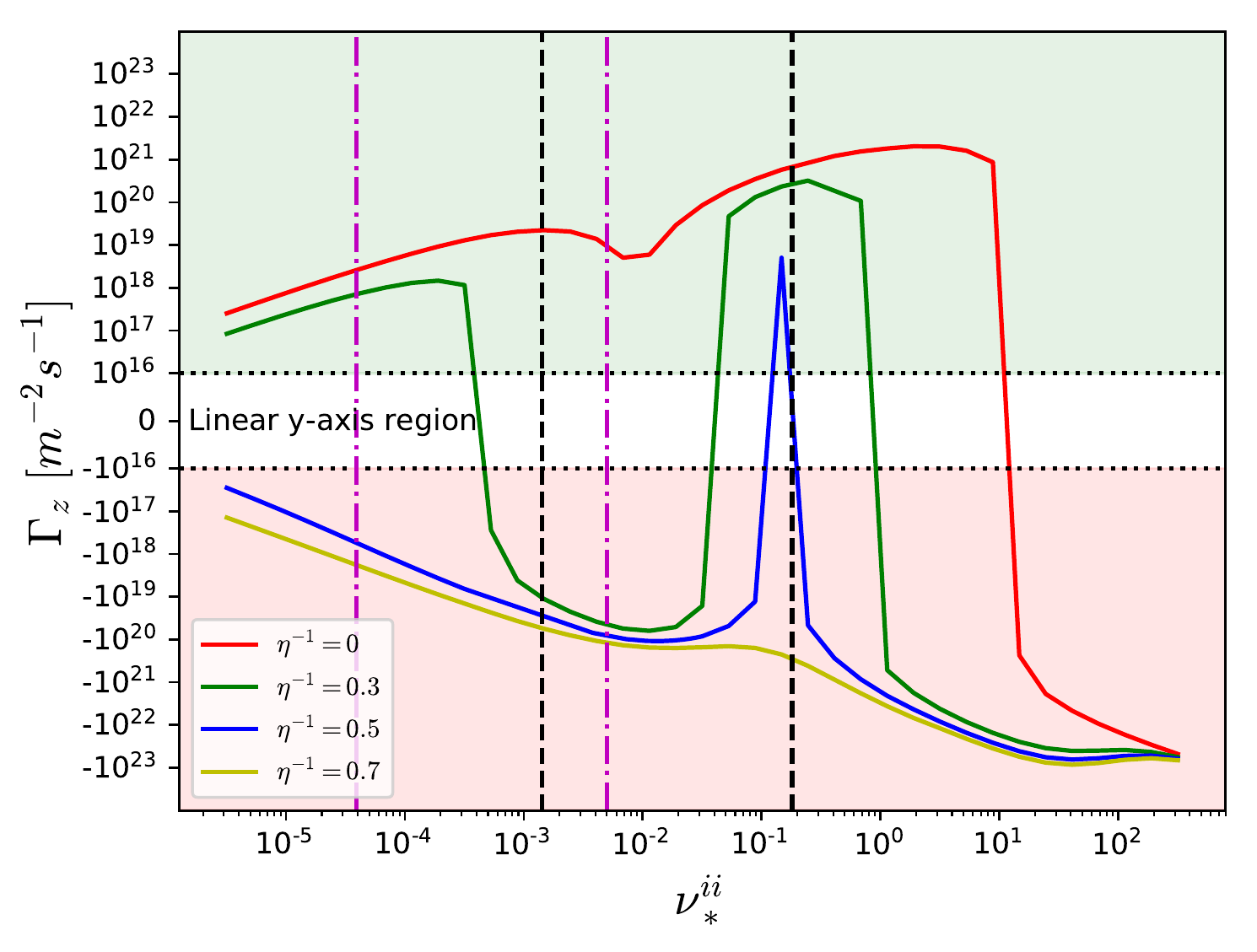}
\caption{(Color online) The impurity particle flux for $C^{6+}$ is plotted in an axisymmetric geometry as a function of the normalized ion-ion collisionality $\nu_*^{ii}\equiv \nu R/v_{ti}$, where $\epsilon=r_N(a/R)$. In this plot, $r_N=0.25$, and $a/R$=0.16. The vertical lines signify the transitions between collisionality regimes for $\nu_*^{ii}$ (black dashed) and $\nu_*^{zz}$ (magenta dotted). The transitions are as follows: $\nu_*^{ii}=1.41\cdot10^{-3}$ (main ion banana-plateau), $\nu_*^{ii}=0.18$ (main ion plateau-Pfirsch-Schl\"uter), $\nu_*^{ii}=3.93\cdot 10^{-5}$ (impurity banana-plateau), and $\nu_*^{ii}=5.03\cdot10^{-3}$ (impurity plateau-Pfirsch-Schl\"uter).
The upper, green-shaded region denotes positive $\Gamma_z$ (impurity screening).
The lower, red-shaded region corresponds to negative $\Gamma_z$ (impurity accumulation).}
\label{fig:axi_coll}
\end{figure}

If it is assumed that both density and temperature profiles are peaked, then achieving a positive (outward) $\Gamma_z$ is based on two properties: $Z_{eff}$, and more importantly, the ratio $\eta^{-1}\equiv d\ln(n)/d\ln(T)$ .
In the absence of a density gradient, the term in square brackets in \ref{eq:gamma_axi} will always be positive and lead to temperature screening.
However, based on the $\alpha$, there is some critical $\eta_c^{-1}(\alpha)\sim 0.4$ where $\eta^{-1}>\eta_c^{-1}$ will always lead to impurity accumulation.
This effect can be seen in Figure \ref{fig:axi_coll}, where SFINCS has been used to calculate the impurity particle flux for $C^{6+}$ over a range of collisionalities for an axisymmetric geometry model $B(\theta)=B_0(1+\epsilon\cos\theta)$, with $\iota=0.689$ and $\alpha=1$.
It is also clear that temperature screening is only accessible up to some maximum collisionality, regardless of $\eta^{-1}$.
However, such high collisionalities put not only impurities, but also bulk ions, into the Pfirsch-Schl\"{u}ter regime, which is generally too collisional to be relevant in the core of reactor scenarios.  

\bibliographystyle{jpp}

\bibliography{jpp-instructions}

\end{document}